\documentclass[aps, prd, reprint,showpacs,  preprintnumbers,amsmath,amssymb,superscriptaddress,  nofootinbib]{revtex4-1}

\usepackage{amssymb, amsmath, braket, nicefrac, graphicx} 
\usepackage[tight]{subfigure}
\usepackage{relsize}
\usepackage{lineno}
\usepackage{setspace}

\usepackage{ esint }
\usepackage{chngpage}
\usepackage{feynmf}
\usepackage{epsfig}
\usepackage{multirow}
\usepackage{rotating}
\usepackage{float}
\usepackage{color}
\usepackage{xcolor}
\usepackage{appendix}
\usepackage{ amstext }
\usepackage{ gensymb }
\usepackage{ textcomp }
\usepackage{ wasysym }

\usepackage{upgreek}

\usepackage{pstricks}

\newcommand{\etapi}{\mbox{$\eta_{\small{\uppi}}$}}

\newcommand{\numu}{\mbox{$\nu_{\mu}$}}                   
\newcommand{\nue}{\mbox{$\nu_{e}$}}                      

\newcommand{\CCne}{\mbox{CC-$\nu_{e}$}}                              
\newcommand{\CCnm}{\mbox{CC-$\nu_{\mu}$}}                            

\newcommand{\anu}{\ensuremath{\bar{\nu}}}

\newcommand{\anumu}{\ensuremath{\bar{\nu}_{\mu}}}

\newcommand{\piz}{\mbox{$\uppi^{0}$}} 
\newcommand{\simgt}{\,\hbox{\lower0.6ex\hbox{$\sim$}\llap{\raise0.6ex\hbox{$>$}}}\,}
\newcommand{\simlt}{\,\hbox{\lower0.6ex\hbox{$\sim$}\llap{\raise0.6ex\hbox{$<$}}}\,}
\definecolor{maroon}{RGB}{162,10,10}

\usepackage{hyperref}
\usepackage[all]{hypcap}
\hypersetup{
    colorlinks,
    citecolor=black,
    filecolor=black,
    linkcolor=black,
    urlcolor=black
}

\makeatletter
\renewenvironment{table}
  {\def\@captype{table}}
  {}

\renewenvironment{figure}
  {\def\@captype{figure}}
  {}
\makeatother
\begin{document}
\preprint{FERMILAB-PUB-16-145-ND}
\title{Measurement of single $\uppi^0$ production by coherent neutral-current 
$\nu$ Fe interactions in the MINOS Near Detector}        
\newcommand{\Berkeley}{Lawrence Berkeley National Laboratory, Berkeley, California, 94720 USA}
\newcommand{\Cambridge}{Cavendish Laboratory, University of Cambridge, 
Cambridge CB3 0HE, United Kingdom}
\newcommand{\Cincinnati}{Department of Physics, University of Cincinnati, Cincinnati, Ohio 45221, USA}
\newcommand{\FNAL}{Fermi National Accelerator Laboratory, Batavia, Illinois 60510, USA}
\newcommand{\RAL}{Rutherford Appleton Laboratory, Science and Technology Facilities Council, Didcot, OX11 0QX, United Kingdom}
\newcommand{\UCL}{Department of Physics and Astronomy, University College London, 
London WC1E 6BT, United Kingdom}
\newcommand{\Caltech}{Lauritsen Laboratory, California Institute of Technology, Pasadena, California 91125, USA}
\newcommand{\Alabama}{Department of Physics and Astronomy, University of Alabama, Tuscaloosa, Alabama 35487, USA}
\newcommand{\ANL}{Argonne National Laboratory, Argonne, Illinois 60439, USA}
\newcommand{\Athens}{Department of Physics, University of Athens, GR-15771 Athens, Greece}
\newcommand{\NTUAthens}{Department of Physics, National Tech. University of Athens, GR-15780 Athens, Greece}
\newcommand{\Benedictine}{Physics Department, Benedictine University, Lisle, Illinois 60532, USA}
\newcommand{\BNL}{Brookhaven National Laboratory, Upton, New York 11973, USA}
\newcommand{\CdF}{APC -- Universit\'{e} Paris 7 Denis Diderot, 10, rue Alice Domon et L\'{e}onie Duquet, F-75205 Paris Cedex 13, France}
\newcommand{\Cleveland}{Cleveland Clinic, Cleveland, Ohio 44195, USA}
\newcommand{\Delhi}{Department of Physics \& Astrophysics, University of Delhi, Delhi 110007, India}
\newcommand{\GEHealth}{GE Healthcare, Florence South Carolina 29501, USA}
\newcommand{\Harvard}{Department of Physics, Harvard University, Cambridge, Massachusetts 02138, USA}
\newcommand{\HolyCross}{Holy Cross College, Notre Dame, Indiana 46556, USA}
\newcommand{\Houston}{Department of Physics, University of Houston, Houston, Texas 77204, USA}
\newcommand{\IIT}{Department of Physics, Illinois Institute of Technology, Chicago, Illinois 60616, USA}
\newcommand{\Iowa}{Department of Physics and Astronomy, Iowa State University, Ames, Iowa 50011 USA}
\newcommand{\Indiana}{Indiana University, Bloomington, Indiana 47405, USA}
\newcommand{\ITEP}{High Energy Experimental Physics Department, ITEP, B. Cheremushkinskaya, 25, 117218 Moscow, Russia}
\newcommand{\JMU}{Physics Department, James Madison University, Harrisonburg, Virginia 22807, USA}
\newcommand{\LASL}{Nuclear Nonproliferation Division, Threat Reduction Directorate, Los Alamos National Laboratory, Los Alamos, New Mexico 87545, USA}
\newcommand{\Lebedev}{Nuclear Physics Department, Lebedev Physical Institute, Leninsky Prospect 53, 119991 Moscow, Russia}
\newcommand{\Lancaster}{Lancaster University, Lancaster, LA1 4YB, UK}
\newcommand{\LLL}{Lawrence Livermore National Laboratory, Livermore, California 94550, USA}
\newcommand{\LosAlamos}{Los Alamos National Laboratory, Los Alamos, New Mexico 87545, USA}
\newcommand{\Manchester}{School of Physics and Astronomy, University of Manchester, 
Manchester M13 9PL, United Kingdom}
\newcommand{\MIT}{Lincoln Laboratory, Massachusetts Institute of Technology, Lexington, Massachusetts 02420, USA}
\newcommand{\Minnesota}{University of Minnesota, Minneapolis, Minnesota 55455, USA}
\newcommand{\Crookston}{Math, Science and Technology Department, University of Minnesota -- Crookston, Crookston, Minnesota 56716, USA}
\newcommand{\Duluth}{Department of Physics, University of Minnesota Duluth, Duluth, Minnesota 55812, USA}
\newcommand{\Ohio}{Center for Cosmology and Astro Particle Physics, Ohio State University, Columbus, Ohio 43210 USA}
\newcommand{\Otterbein}{Otterbein University, Westerville, Ohio 43081, USA}
\newcommand{\Oxford}{Subdepartment of Particle Physics, University of Oxford, Oxford OX1 3RH, United Kingdom}
\newcommand{\PennState}{Department of Physics, Pennsylvania State University, State College, Pennsylvania 16802, USA}
\newcommand{\PennU}{Department of Physics and Astronomy, University of Pennsylvania, Philadelphia, Pennsylvania 19104, USA}
\newcommand{\Pittsburgh}{Department of Physics and Astronomy, University of Pittsburgh, Pittsburgh, Pennsylvania 15260, USA}
\newcommand{\IHEP}{Institute for High Energy Physics, Protvino, Moscow Region RU-140284, Russia}
\newcommand{\Rochester}{Department of Physics and Astronomy, University of Rochester, New York 14627 USA}
\newcommand{\RoyalH}{Physics Department, Royal Holloway, University of London, Egham, Surrey, TW20 0EX, United Kingdom}
\newcommand{\Carolina}{Department of Physics and Astronomy, University of South Carolina, Columbia, South Carolina 29208, USA}
\newcommand{\SDakota}{South Dakota School of Mines and Technology, Rapid City, South Dakota 57701, USA}
\newcommand{\SLAC}{Stanford Linear Accelerator Center, Stanford, California 94309, USA}
\newcommand{\Stanford}{Department of Physics, Stanford University, Stanford, California 94305, USA}
\newcommand{\StJohnFisher}{Physics Department, St. John Fisher College, Rochester, New York 14618 USA}
\newcommand{\Sussex}{Department of Physics and Astronomy, University of Sussex, Falmer, Brighton BN1 9QH, United Kingdom}
\newcommand{\TexasAM}{Physics Department, Texas A\&M University, College Station, Texas 77843, USA}
\newcommand{\Texas}{Department of Physics, University of Texas at Austin, 
Austin, Texas 78712, USA}
\newcommand{\TechX}{Tech-X Corporation, Boulder, Colorado 80303, USA}
\newcommand{\Tufts}{Physics Department, Tufts University, Medford, Massachusetts 02155, USA}
\newcommand{\UNICAMP}{Universidade Estadual de Campinas, IFGW, CP 6165, 13083-970, Campinas, SP, Brazil}
\newcommand{\UFG}{Instituto de F\'{i}sica, 
Universidade Federal de Goi\'{a}s, 74690-900, Goi\^{a}nia, GO, Brazil}
\newcommand{\USP}{Instituto de F\'{i}sica, Universidade de S\~{a}o Paulo,  CP 66318, 05315-970, S\~{a}o Paulo, SP, Brazil}
\newcommand{\Warsaw}{Department of Physics, University of Warsaw, 
PL-02-093 Warsaw, Poland}
\newcommand{\Washington}{Physics Department, Western Washington University, Bellingham, Washington 98225, USA}
\newcommand{\WandM}{Department of Physics, College of William \& Mary, Williamsburg, Virginia 23187, USA}
\newcommand{\Wisconsin}{Physics Department, University of Wisconsin, Madison, Wisconsin 53706, USA}
\newcommand{\deceased}{Deceased.}

\affiliation{\ANL}
\affiliation{\Athens}
\affiliation{\BNL}
\affiliation{\Caltech}
\affiliation{\Cambridge}
\affiliation{\UNICAMP}
\affiliation{\Cincinnati}
\affiliation{\FNAL}
\affiliation{\UFG}
\affiliation{\Harvard}
\affiliation{\HolyCross}
\affiliation{\Houston}
\affiliation{\IIT}
\affiliation{\Indiana}
\affiliation{\Iowa}
\affiliation{\Lancaster}
\affiliation{\UCL}
\affiliation{\Manchester}
\affiliation{\Minnesota}
\affiliation{\Duluth}
\affiliation{\Otterbein}
\affiliation{\Oxford}
\affiliation{\Pittsburgh}
\affiliation{\RAL}
\affiliation{\USP}
\affiliation{\Carolina}
\affiliation{\Stanford}
\affiliation{\Sussex}
\affiliation{\TexasAM}
\affiliation{\Texas}
\affiliation{\Tufts}
\affiliation{\Warsaw}
\affiliation{\WandM}

\author{P.~Adamson}
\affiliation{\FNAL}


\author{I.~Anghel}
\affiliation{\Iowa}
\affiliation{\ANL}



\author{A.~Aurisano}
\affiliation{\Cincinnati}









\author{G.~Barr}
\affiliation{\Oxford}









\author{M.~Bishai}
\affiliation{\BNL}

\author{A.~Blake}
\affiliation{\Cambridge}
\affiliation{\Lancaster}


\author{G.~J.~Bock}
\affiliation{\FNAL}


\author{D.~Bogert}
\affiliation{\FNAL}




\author{S.~V.~Cao}
\affiliation{\Texas}

\author{T.~J.~Carroll}
\affiliation{\Texas}

% ...

\collaboration{The MINOS Collaboration}
\noaffiliation

\date{\today}
\pacs{12.15.Mm, 13.15+g, 25.30.Pt}

\begin{abstract}

Forward single $\uppi^0$ production by coherent neutral-current interactions, 
$\nu\,\mathcal{A} \to \nu\,\mathcal{A}\,\uppi^0$, is investigated using 
a 2.8$\times 10^{20}$ protons-on-target exposure of the MINOS Near Detector.
For single-shower topologies, the event distribution in production angle
exhibits a clear excess above the estimated background at very forward angles for visible energy in the range~1-8\,GeV.
Cross sections are obtained for the detector medium comprised of 80\% iron and 20\% carbon nuclei 
with $\langle \mathcal{A} \rangle = 48$, the highest-$\langle \mathcal{A} \rangle$ target 
used to date in the study of this coherent reaction. 
The total cross section for coherent neutral-current single-$\uppi^0$ production initiated 
by the $\numu$ flux of the NuMI low-energy beam with mean (mode) $E_{\nu}$ of 4.9 GeV (3.0 GeV),
is $77.6\pm5.0\,(\text{stat}) ^{+15.0}_{-16.8}\,(\text{syst})\times10^{-40}\,\text{cm}^2~\text{per nucleus}$. 
The results are in good agreement with predictions of the Berger-Sehgal model.

\end{abstract}

\maketitle

\section{Introduction} \vspace{-9pt}
\label{sec:Intro}

\subsection{$\nu$ NC($\uppi^0$) coherent scattering}

It is well established that single pions
can be produced when a neutrino or antineutrino scatters coherently
from a target nucleus~\cite{ref:Paschos-Schalla-2009}.    These
interactions can proceed either as neutral-current (NC) or charged-current (CC)
processes in which the pion electric charge coincides with 
that of the Z$^{0}$ or W$^{\pm}$ vector boson emitted by the leptonic current. 
Recent investigations, both experimental~\cite{ref:K2K_2, ref:SciBooNE-CC, ref:NOMAD, ref:MiniBooNE, ref:SciBooNE} and 
theoretical~\cite{ref:Paschos, ref:RS_paper_2, ref:BS, ref:Hernandez-2009, 
ref:Singh-2006, ref:Alvarez-Ruso-2009, ref:Martini-2009, ref:Amaro-2009, ref:Nakamura-2010, ref:Hernandez-2010},
have devoted attention to neutrino-induced 
NC coherent production of single $\uppi^0$ mesons:
\begin{equation}
 \nu (\anu) + \mathcal{A} \rightarrow \nu (\anu) + \mathcal{A} + \uppi^{0} \bf{ .}
 \label{eq:nccoh_reaction} 
\end{equation}

Reaction (\ref{eq:nccoh_reaction}) is of theoretical interest
as a process dominated by the divergence of the 
isovector axial-vector neutral current and therefore amenable
to calculation using the Partially Conserved Axial-Vector Current (PCAC) hypothesis
 and Adler's theorem~\cite{ref:Adler-theorem}.   
The phenomenological model of Rein and Sehgal~\cite{ref:RS_paper_1}  
 invokes Adler's theorem to express the coherent cross section 
 in terms of the $\uppi$-nucleon scattering cross section.  
 The original Rein-Sehgal model characterized
coherent scattering at incident energies $E_\nu > 3$\,GeV, 
and served as a framework for development of other PCAC-based models 
of coherent $\uppi^{0}$ production~\cite{ref:RS_paper_2, ref:BS, ref:Paschos, ref:Hernandez-2009}.  
In particular, the Berger-Sehgal model~\cite{ref:BS} used in the present work
improves upon Rein-Sehgal by using $\uppi$-carbon scattering data 
rather than $\uppi$-nucleon data as the basis for extrapolation.

An alternative class of models, appropriate for sub-GeV to few-GeV neutrino scattering, has also received considerable
attention~\cite{ref:Singh-2006, ref:Alvarez-Ruso-2009, 
ref:Martini-2009, ref:Amaro-2009, ref:Nakamura-2010, ref:Hernandez-2010}.   
In these ``dynamical models" the amplitudes for
various neutrino-nucleon reactions yielding the single pion final state are added coherently over the 
nucleus.  Within the past decade the theoretical descriptions of coherent NC $\uppi^0$ production for $E_{\nu}$ below
a few GeV have achieved a level of detail previously unavailable~\cite{ref:Boyd-2009}.    

Experimental investigations of coherent NC($\uppi^0$) production to date have been
limited to scattering on targets with an average nucleon number, $\langle \mathcal{A} \rangle$,
in the range $\langle \mathcal{A} \rangle \leq 30$ (see Table~~\ref{tab:prev_exp}).  
In the study reported here, the cross section for Reaction\,\eqref{eq:nccoh_reaction} is measured using
a high statistics sample of neutrino interactions recorded by the MINOS Near Detector~\cite{ref:NIM, ref:CCinclusive}.
The Near Detector consists of iron plates interleaved with plastic scintillator, 
yielding an average nucleon number of 48.   Thus the MINOS measurement probes the coherent
Reaction\,\eqref{eq:nccoh_reaction} using a target with $\langle \mathcal{A} \rangle$ 
distinctly higher than utilized previously, as detailed in Sec.\,\ref{subsec:Prev-meas}.

\subsection{Reaction phenomenology}

\begin{figure}
    \centering
 \scalebox{0.30}{\includegraphics{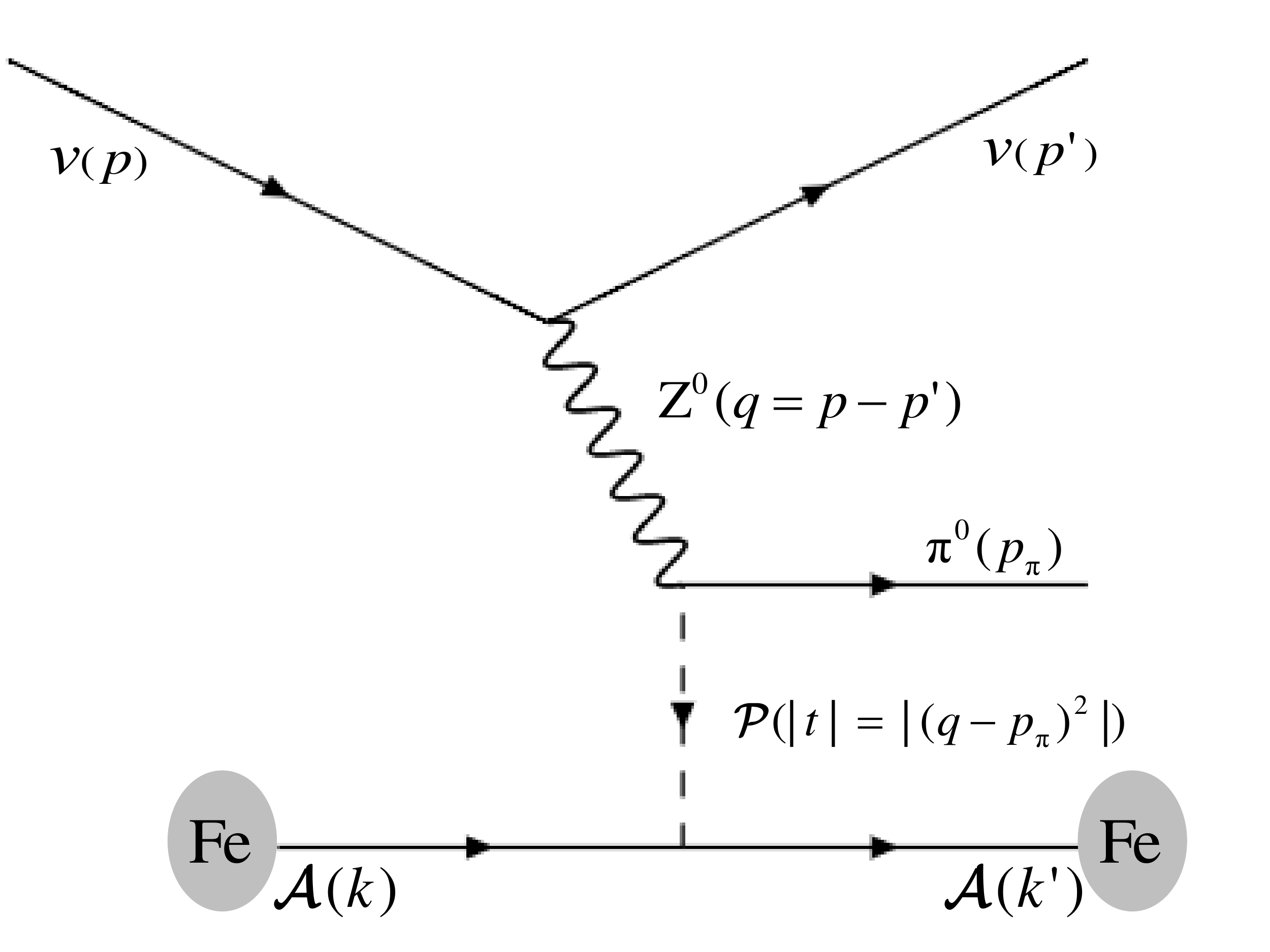}}   
 \caption{Mechanism for neutrino-nucleus NC($\uppi^0$) coherent scattering.   
  The Z$^{0}$ boson initiates virtual $\uppi^0$ elastic scattering
  with exchange of a pomeron-like quantum ($\mathcal{P}$) 
  which transfers four-momentum squared $|t|$ to the nucleus.}
  \label{Fig01}
\end{figure}

In coherent scattering no quantum numbers are transferred to the target nucleus, and
the square of the four-momentum transfer to the nucleus, 
$|t| = |(q - p_{\pi})^2|$, is very small.   
Figure~\ref{Fig01} depicts the amplitude proposed
by Rein and Sehgal to describe coherent NC($\uppi^0$) production 
in the limit $Q^{2} \equiv -q^{2} = -(p - p')^{2} \rightarrow 0$ 
where both the Conserved Vector Current (CVC) and the PCAC hypotheses apply.   
The differential cross section away from $Q^2$ = 0 can be estimated using the hadron 
dominance model~\cite{ref:HDM, ref:K-M}.   In the Rein-Sehgal and Berger-Sehgal models this
is accomplished using a dipole term of the form $(M_A^2/(M_A^2+Q^2))^2$. 

\smallskip
 
The four-momentum of the final state lepton is not measurable in NC reactions
and so $|t|$ cannot be ascertained.  However, the $Q^2$ dependence can be related
to the observable $\etapi$ which is a measure of the 
momentum transverse to the incident beam~\cite{ref:MiniBooNE, ref:Amaro-2009}:   
\begin{equation}
 \etapi = E_{vis} \times (1 - \cos\theta_{shw})\bf{ .}  
 \label{eq:eta-pi-variable} 
\end{equation}
Here, $E_{vis}$ is the visible energy of the gamma conversions resulting from
$\uppi^0$ decay and $\theta_{shw}$ is the angle of the electromagnetic shower
with respect to the beam direction.
Distributions of $\etapi$ for coherent NC($\uppi^0$) events exhibit
a distinctive peak at low values (see Sec.\,\ref{sec:CohEvts}). 
However it is $\cos\theta_{shw}$ and $E_{vis}$, rather than $\etapi$, that serve as the basic
observables for the MINOS measurement.    The analysis uses event distributions in these two
variables to construct its background model and to extract the signal.

\subsection{Previous measurements}
\label{subsec:Prev-meas}

The first evidence for coherent neutrino-nucleus scattering was
obtained by the Aachen-Padova collaboration using 
spark chambers constructed of aluminum plates~\cite{ref:AachenPadova}.  
Other coherent scattering measurements were carried out during the 1980s
using neutrino beams with different spectra and different
target nuclei~\cite{ref:Gargamelle, ref:SKAT, ref:CHARM, ref:15BC}.
More recently, the NOMAD and SciBooNE experiments have
measured the coherent NC($\uppi^0$) cross section on carbon~\cite{ref:NOMAD, ref:SciBooNE}.   
The MiniBooNE experiment has determined the ratio, $f_{coh}$, 
of NC($\uppi^0$) coherent to coherent-and-resonant production  on carbon:
$f_{coh} = 0.195 \pm  0.011$(stat)$\pm 0.025$(sys) \cite{ref:MiniBooNE}.
The latter measurements, together with searches for 
coherent CC($\uppi^{\pm}$) scattering  by K2K~\cite{ref:K2K_2}
and SciBooNE~\cite{ref:SciBooNE-CC}, stimulated further
theoretical work~\cite{ref:Wascko-review}.  
The coherent NC($\uppi^0$) cross sections for all previous experiments 
are summarized in Table~\ref{tab:prev_exp}. 

\smallskip
\begin{table}
\begin{adjustwidth}{-1.0in}{-1.0in}
\centering
\scalebox{0.70}{%
\begin{tabular}{|c||c|c|c|c|c|c|}
\hline
\multirow{4}{*}{\bf{Experiment}} &                  &               &                       & \multicolumn{3}{|c|}{\bf{{Cross Section per nucleus}}} \\
\cline{5-7}
                           & $<E_\nu>$    & Target $\langle \mathcal{A} \rangle$      & $E^{min}_{\uppi^0}$     & $\upsigma$                 & $\upsigma$/$\upsigma_{\mathrm{R-S}}$ & \multirow{2}{*}{$\upsigma_{\mathrm{B-S}}$}  \\
                           &                        &                                    &                                      & $\nu$($\bar{\nu}$) & $\nu$($\bar{\nu}$)                               &  \\
\cline{2-7}\rule{0pt}{2.8ex}
                           & [GeV]            & [u]         & [GeV]                 & $10^{-40} {\rm cm}^2$  & -         & $10^{-40} {\rm cm}^2$ \\
\hline
\hline
Aachen-                        & \multirow{2}{*}{2} & Aluminum & \multirow{2}{*}{0.18} & 29$\pm$10  & - & \multirow{2}{*}{31}\\
Padova~\cite{ref:AachenPadova} &                    & 27       &                       & (25$\pm$7) & - & \\
\hline
Gargamelle & \multirow{2}{*}{3.5} & Freon           & \multirow{2}{*}{0.2} & 31$\pm$20   & - &\multirow{2}{*}{45}\\
\cite{ref:Gargamelle}                                                  &                    & CF$_{3}$Br - 30 &                      & (45$\pm$24) & - & \\
\hline
CHARM & 31 & Marble          & \multirow{2}{*}{6.0} & 96$\pm$42   & - & \multirow{2}{*}{82}\\
\cite{ref:CHARM}                                        & 24 & CaCO$_{3}$ - 20 &                      & (79$\pm$26) & - & \\
\hline
SKAT& \multirow{2}{*}{7} & Freon           & \multirow{2}{*}{0.2} & \multirow{2}{*}{52$\pm$19} & - & \multirow{2}{*}{62}\\
\cite{ref:SKAT}                                      &                    & CF$_{3}$Br - 30 &                      &                            & - & \\
\hline
15' BC & \multirow{2}{*}{20} & Neon       & \multirow{2}{*}{2.0} & - &\multirow{2}{*}{0.98$\pm$0.24} & \multirow{2}{*}{71}\\
\cite{ref:15BC}                                       &                     &NeH$_{2}$ - 20&                    & - &                               & \\
\hline
NOMAD & \multirow{2}{*}{24.8} & Carbon+ & \multirow{2}{*}{0.5} & \multirow{2}{*}{72.6$\pm$10.6} & - & \multirow{2}{*}{53}\\
\cite{ref:NOMAD}                                        &                       & 12.8    &                      &                                & - & \\
\hline
SciBooNE & \multirow{2}{*}{0.8} & Polystyrene & \multirow{2}{*}{0.0} & - & \multirow{2}{*}{0.96$\pm$0.20} & \multirow{2}{*}{9}\\
\cite{ref:SciBooNE}                                               &                      & C$_{8}$H$_{8}$ - 12 &              & - &                                & \\
\hline
\end{tabular}}
\end{adjustwidth}
\caption{Previous cross section measurements for Reaction\,\eqref{eq:nccoh_reaction}.   
Cross sections as directly reported are displayed in column 5;  values reported as ratios to
Rein-Sehgal $\upsigma_{\mathrm{R-S}}$ are listed in column 6.   Cross sections obtained using
$\anu$ beams are given in parentheses.  Column 7 lists corresponding 
predictions ($\upsigma_{\mathrm{B-S}}$) from the Berger-Sehgal model.}
\label{tab:prev_exp}
\end{table}

\smallskip
Recently, measurements of  charged-current coherent scattering cross sections on carbon 
and on argon, $\numu\,(\anumu) + \mathcal{A} \rightarrow \mu^{\mp} + \mathcal{A} + \uppi^{\pm}$,
have been reported by MINER$\nu$A~\cite{ref:CC-Coh-Minerva} 
and by ArgoNeuT~\cite{ref:CC-Coh-Argon} respectively.   The neutrino fluxes for these measurements, obtained
with operation of the NuMI beam in low energy mode, are similar to the neutrino flux used for the present study.
For neutrino-nucleus scattering at $E_{\nu} > $ 3 GeV, the PCAC models predict
the final-state pion kinematics for coherent NC($\uppi^{0}$) scattering to be very
similar to the kinematics observed in coherent CC($\uppi^{\pm}$) scattering.  
Consequently the distributions reported for the full range of $E_{\pi}$ from
CC($\uppi^{\pm}$) coherent scattering~\cite{ref:CC-Coh-Minerva}
provide guidance for estimation of  the coherently-produced $\uppi^{0}$ rate 
below the MINOS threshold for electromagnetic (EM) shower detection.

\section{Analysis Overview}
\label{sec:IntroAnal}
       
Measurement of the NC($\uppi^0$) coherent scattering cross section
requires that this rare reaction, predicted to constitute about $0.2$\% of all neutrino interactions in the exposure, be detected
amidst a copious background of neutrino reactions having topologies that are dominated by an EM shower.
The background is mostly composed of NC reactions wherein an incoherently
produced, energetic $\uppi^0$ dominates the final-state.   Backgrounds also arise from
energetic $\uppi^0$ initiated by CC $\numu$ interactions with large fractional energy transfer to the hadronic system,
and from quasielastic-like CC $\nue$ interactions.
 
This analysis uses a reference Monte Carlo (MC) event sample simulated using 
the NEUGEN3 event generator~\cite{ref:neugen3} and other codes of 
the standard MINOS software framework~\cite{ref:minos-QE}.
The reference MC sample includes NC($\uppi^0$) coherent scattering 
generated according to the Berger-Sehgal model. 

Candidate events are isolated by requiring containment within the fiducial volume, 
absence of charged particle tracks, and visible energy 
sufficient to reconstruct an EM shower with $E_{vis} >$ 1.0\,GeV.    
Further background reduction is achieved 
by distinguishing electromagnetic from hadronic-shower behavior
using a multivariate analysis classification algorithm known as
Support Vector Machines~\cite{ref:Statistical-Learning, ref:LibSVM}.   

Subsamples of the selected MC sample are organized and handled as binned event distributions that
are functions of the kinematic variables $\cos\theta_{shw}$ and $E_{vis}$.
An event distribution of this kind constitutes a ``template" 
over the plane of $\cos\theta_{shw}$-versus-$E_{vis}$ (discussed in Sec.\,\ref{sec:EvtSel}).   
Each of the different background reaction categories 
is embodied by its template distribution. 
These subsample templates extend over the signal region (defined by a relatively high signal-to-background ratio)
and over the sidebands (kinematic regions adjacent 
or close to the signal region with low predicted signal content).

The background templates are constrained by fitting to data events
in the sidebands.  The fit adjusts the normalizations and shapes
of the background templates using normalization fit parameters plus
two systematic parameters;   the latter account for the effects
of specific sources of uncertainty capable of generating template shape distortions. 
Fitting to sidebands is restricted to regions that, according to the MC, have signal purity less than $5\%$,
since optimization studies showed this cut to minimize the total uncertainty propagated to the measurement.
The ensemble of templates, fit to the sidebands, define a background model that also
extends over the signal region of the $\cos\theta_{shw}$-vs-$E_{vis}$ plane.   

The formalism used to subtract background from data 
in the signal region is discussed in Secs.\,\ref{sec:BgrByFitting} and \ref{sec:ExtractSig}.
The delineation, evaluation, and method of treatment 
of systematic uncertainties are presented in Sec.\,\ref{sec:SystUnc}.
At this point the foundation is set for fitting the background model to the
data sidebands.   Results of this fit are given in Sec.\,\ref{sec:FitSideData}, and
the background rate over the signal region is thereby established.  
The subtraction of the background from the data in the signal region yields the 
measured number of NC($\uppi^0$) coherent scattering events (Sec.\,\ref{sec:Xsec}), enabling the scattering
cross sections to be determined (Sec.\,\ref{sec:tot_xsec}).
Section~\ref{sec:Results} discusses the MINOS
cross sections in the context of previously reported
NC($\uppi^0$) coherent scattering measurements and 
summarizes the observational results of this work.

Data blinding protocols were used throughout the development of the analysis.
Data bins for which the signal purity was predicted by the MC simulation to exceed
$20\%$ were always masked.   Additionally, protocols were followed that forbade data versus MC
comparisons and fits involving the data sidebands until all work to establish the fit procedure was completed.

\subsection{Flux-averaged cross section measurement}
\label{subsec:Xsec-measment}

For coherent NC($\uppi^0$) events, the visible energy of the final-state $\uppi^0$ 
is only a fraction of the incident neutrino energy, $E_{\nu}$.
Extraction of the reaction cross section as a function of $E_{\upnu}$ is therefore problematic.   
Nevertheless, a flux-averaged cross section, $\langle\upsigma\rangle$,  
representative of a designated $E_{\nu}$ range can be measured.
Let $\mathcal{N}_T$ denote the number of target nuclei in the Near Detector fiducial volume.
The total neutrino flux for the experiment is $\mathcal{N}_{\text p} \times \Phi$, where 
$\mathcal{N}_{\text p}$ is the total number of protons on target (POT) and $\Phi$ is the
integral over $E_{\nu}$ of the flux spectrum per POT at the front surface of the fiducial volume, $\phi(E_{\nu})$:
$\Phi = \int \phi(E_{\nu}){\text d}E_{\nu}$.  
The number of reactions after correction for detection inefficiencies, $N^{Coh}$, is given by
\begin{eqnarray}
N^{Coh} = ~\mathcal{N}_{T}\, \mathcal{N}_{\text p} \int \upsigma(E_{\nu}) \, \phi(E_{\nu})\, {\text d}E_{\nu} ~,
\label{eq:xsec2}
\end{eqnarray}
so that
\begin{eqnarray}
\langle\upsigma\rangle = \frac{N^{Coh}}{ \mathcal{N}_{T}\,\mathcal{N}_{\text p}\,\Phi}~ \bf{ .}
\label{eq:xsec3}
\end{eqnarray}
The constants $\mathcal{N}_{T}$, $\mathcal{N}_{\text p}$, and $\Phi$ 
are determined by the experimental setup and running conditions.
The fully corrected count of signal events, $N^{Coh}$, 
effectively measures the flux-averaged cross section.

\section{Beam, Detector, Data Exposure} \vspace{-9pt}
\label{sec:B-D-DE}

\subsection{Neutrino beam and Near Detector}
\label{sec:MINOSND}

During the running of the MINOS experiment, the 
Neutrinos at the Main Injector (NuMI) beam~\cite{ref:NuMI-beam} used a primary beam of
120\,GeV protons delivered by the Main Injector in 10\,$\upmu$s spills 
every $2.2\,\text{s}$.    The protons were directed onto a graphite target, producing 
large numbers of hadronic particles.  The produced hadrons 
traversed two magnetic focusing horns whose current
polarity was set to focus positively charged particles (mostly $\uppi^+$ and K$^+$ mesons),
directing them into a 675\,m long cylindrical decay pipe.   Positioned downstream of the decay pipe was 
the hadron absorber, followed by 240\,m of rock to stop the remaining muons.
Along the first 40\,m of rock there were three 
alcoves, each containing a plane of muon monitoring chambers
that measured the muon flux.   

The Near Detector data were obtained using the low-energy (LE) beam
configured with the downstream end of the target inserted 50.4\,cm into the first (most upstream) horn
and with 185\,kA currents in the two horns.   With the LE beam in neutrino
mode, the wide-band neutrino spectrum peaked at 3.0\,GeV and had an average
neutrino energy  $\langle{E_\nu}\rangle$ = 4.9\,GeV.    
The relative rates of CC interactions by incident neutrino type were estimated to be
91.7\% $\nu_{\mu}$, 7.0\% $\bar{\nu}_{\mu}$, 1.0\% $\nu_{e}$, and 
0.3\% $\bar{\nu}_{e}$.   Details concerning beam layout, instrumentation, and
neutrino spectrum are given in Ref.~\cite{ref:NuMI-beam}.

The MINOS Near Detector is a sampling tracking calorimeter of 980 metric tons 
located 1.04 km downstream of the beam target in a cavern 103\,m underground.  The detector is 
composed of interleaved, vertically mounted planes.  
Each plane contains a 2.54\,cm thick steel layer and a 1.0\,cm thick scintillator layer,
providing 1.4 radiation lengths per plane.
The plastic strips of a scintillator plane are  oriented 45$\degree$ from the horizontal, 
with each plane (a ``U-plane" or ``V-plane") rotated 90$\degree$ from the previous plane.   
The detector steel is magnetized with a toroidal field having an average intensity of 1.3\,T.  

The requirements of full containment, isolation from hadronic (non-EM) showers, 
and optimal reconstruction for candidate EM showers 
are the same here as for the MINOS $\nue$ appearance measurements, consequently the same fiducial
volume within the Near Detector is used~\cite{ref:T-Yang_Thesis, ref:Boehm_Thesis, ref:Ochoa_Thesis}.
The fiducial volume is a cylinder of 0.8\,m radius and of 4.0\,m length in the beam direction.    
Full descriptions of the scintillator strip configuration, 
event readout, and off-line processing, are given in Refs.~\cite{ref:NIM, ref:CCinclusive}.

The bulk mass of the detector resides in its steel plates.
The scintillator strips and other components account for less than 5\% of the mass.
Uncertainty in the fiducial mass reflects measurement errors
for the widths and mass of the steel plates; 
it is estimated to be $\pm$0.4\% \cite{ref:NIM}.  
There are $(3.57\pm 0.01)\times$10$^{29}$ nuclei within the fiducial 
volume of which $\sim$80\% are iron nuclei and $\sim$20\% are
carbon nuclei, yielding an average atomic mass of $\langle \mathcal{A} \rangle$ $= 48$\,u.   

The electromagnetic and hadronic shower
 energies are determined using calorimetry.    The absolute
energy scale for the Near Detector EM shower response has been determined to within
$\pm5.6\%$~\cite{ref:NIM, ref:VahleThesis, ref:KordThesis}.

\begin{figure}
\begin{adjustwidth}{-1.0in}{-1.0in}
\centering
\scalebox{0.88}{\includegraphics{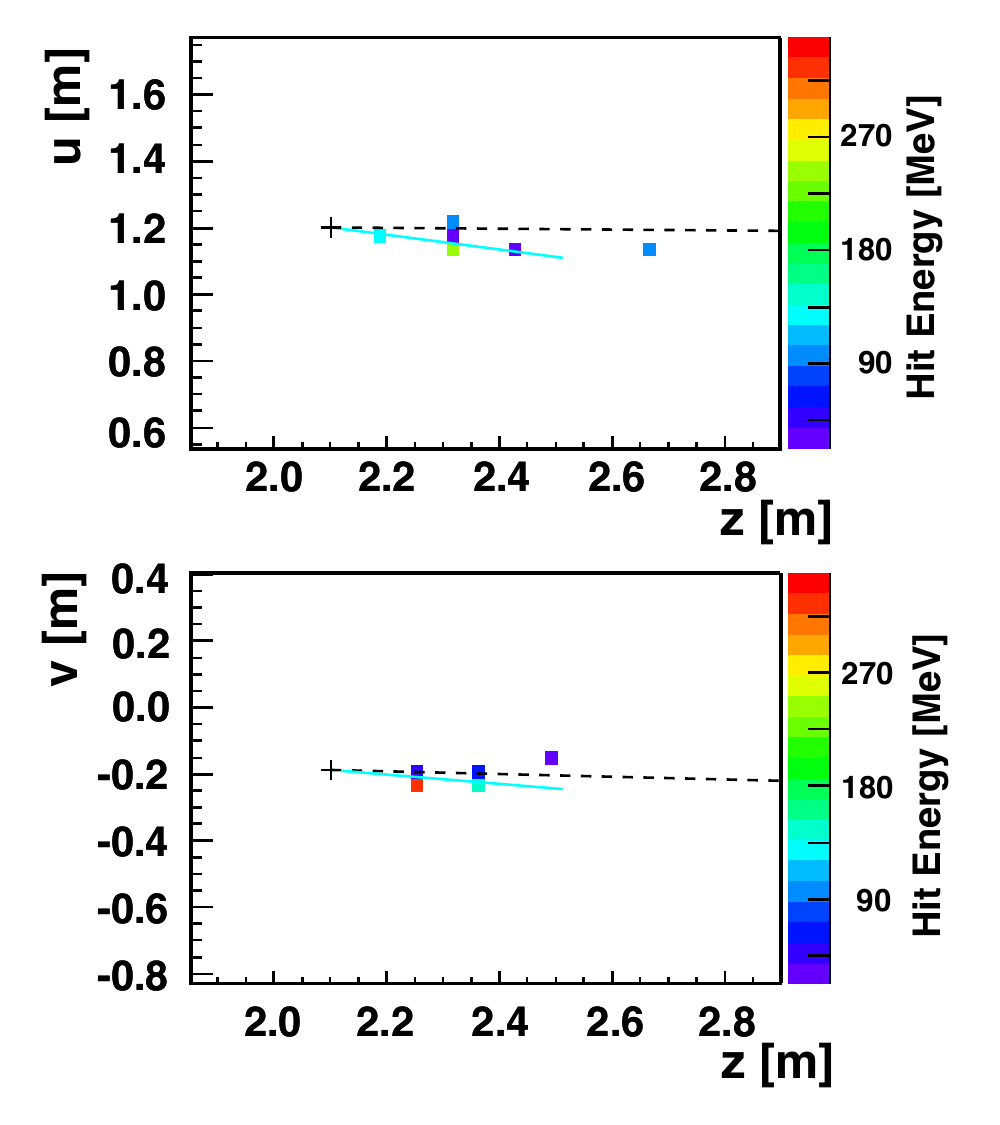}}
\end{adjustwidth}
\caption{Simulation of coherent 
 $\nu\,+\,\text{Fe} \rightarrow \nu\, +\,\text{Fe}\,+\,\uppi^{0}$ in the Near Detector,
 showing the locations of hits projected in U-view (upper plot) and V-view (lower plot).  
 The rightmost scale gives the energy deposition in scintillator.
 Dashed black and solid cyan lines show trajectories of the final state neutrino and $\uppi^0$ respectively. }
\label{Fig02}
\end{figure}

\subsection{Data exposure and neutrino flux}
\label{sec:expo}

The data are obtained from a total exposure of $2.8\times10^{20}$ POT, from
MINOS runs of May 2005 through July 2007. 
The POT count is accurate to within 1.0\%~\cite{ref:POT_Error}. 
The data set was estimated at the outset to be enough to ensure that the 
measurement would be limited by systematics rather than statistics. 
 The final results vindicate that estimate; the statistical uncertainties 
 are generally smaller than systematic uncertainties.
 
A determination of the LE beam $\nu_{\mu}$ flux for the data used in this work
was obtained as part of the MINOS measurement 
of the inclusive CC-$\numu$ cross section~\cite{ref:CCinclusive}. 
The determination was based upon analysis of a CC subsample characterized by low energy transfer, $\upnu$,
to the hadronic system.   The data rate in this subsample measures the $\nu_{\mu}$ flux because, in the limit
of low-$\upnu$, the differential cross section $d\upsigma^{\nu}/d\upnu$ approaches 
a constant value independent of $E_{\nu}$~\cite{ref:low-nu-method, ref:D-Bhat_Thesis}.
Binned values for the $\nu_{\mu}$ flux and uncertainties for $3.0< E_{\nu}<50$\,GeV are given in Table II 
of Ref.~\cite{ref:CCinclusive}.

In a separate determination, the muon fluxes downstream of the beam decay pipe were measured at
various target positions and for different horn currents using monitoring chambers deployed in the three rock alcoves.
An {\it ab initio} simulation of the $\nu_{\mu}$ flux was then adjusted to match the muon flux 
observations~\cite{ref:LoiaconoThesis}.   The two determinations gave similar neutrino fluxes for the
the $E_{\nu}$ range above 3.0 GeV where they overlap.   For the analysis of this work,
the more precise $\numu$ flux determination of Ref.~\cite{ref:CCinclusive} is used
for $E_\nu > 3.0$\,GeV, while the $\numu$ flux calculation 
constrained by measured muon fluxes is used for $E_\nu < 3.0$\,GeV. 
The  neutrino flux integrated from 0.0 to 50\,GeV is (2.93$\pm$0.23)\,$\nu /\text{m}^{2}$/$10^4$\,POT. 
The average $E_\nu$ is 4.9\,GeV and the spectral peak is at 3.0\,GeV.  The range of neutrino energies 
about $\langle{E_\nu}\rangle$ that contains 68\% of the flux is
2.4~$\leq E_\nu \leq$~9.0\,GeV.   Based upon the measurements reported in 
Refs.~\cite{ref:CCinclusive} ($E_{\nu} > 3.0$ GeV) and~\cite{ref:LoiaconoThesis} ($E_{\nu} < 3.0$ GeV),
an uncertainty of 7.8$\%$ is assigned to the integrated flux.

\section{Coherent NC($\uppi^0$) Events}
\label{sec:CohEvts}

An example simulation of a NC($\uppi^0$) coherent interaction  
in the Near Detector is shown in Fig.~\ref{Fig02}.
A single $\uppi^0$ meson of energy 1.31\,GeV is produced at a vertex located two 
scintillator planes upstream of the gamma conversions.   The two gamma conversions appear
as a single 1.28\,GeV electromagnetic shower.
In general, electromagnetic showers and hadronic showers of individual events 
can be distinguished using the reconstructed energy deposition patterns.

\begin{figure}
\begin{adjustwidth}{-1.0in}{-1.0in}
\centering
\scalebox{0.46}{\includegraphics[trim = 0mm 1mm 0mm 3mm, clip]{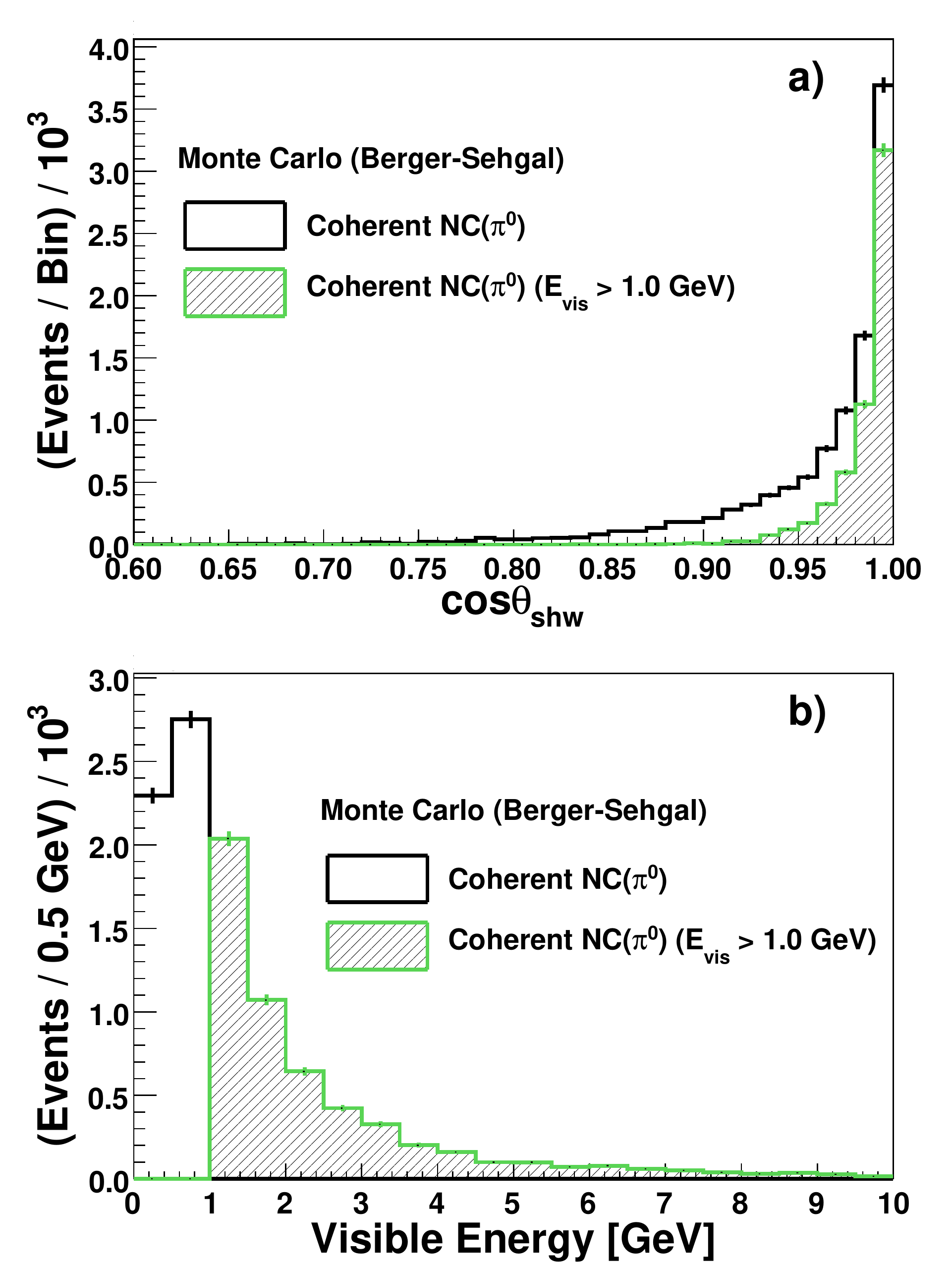}}
\end{adjustwidth}  
\caption{Monte Carlo distributions 
for NC($\uppi^0$) coherent interactions (Berger-Sehgal model) in the Near Detector for 
the LE beam exposure.   (a) Shower-angle cosine of final-state
showers with respect to the $\nu$ beam, and   (b) Event visible energies.
Shaded regions show events with $E_{vis} > 1$\,GeV. }
\label{Fig03}
\end{figure}

\begin{figure}
\begin{adjustwidth}{-1.0in}{-1.0in}
\centering
\scalebox{0.46}{\includegraphics[angle=0, trim = 1mm 1mm 0mm 4mm, clip]{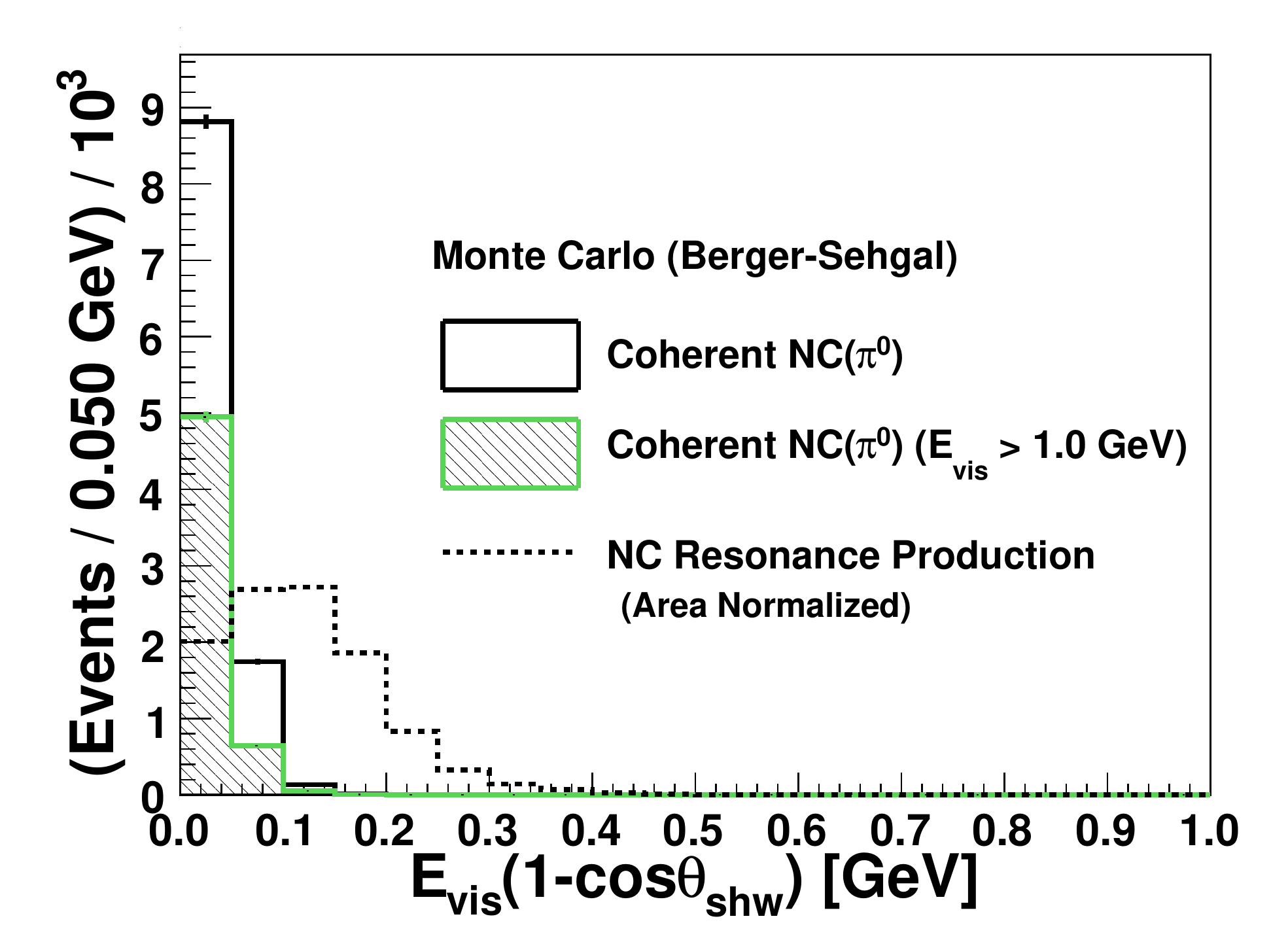}}
\end{adjustwidth}  
\caption{Monte Carlo distribution (solid histogram) of the $\etapi$ variable for 
neutrino NC($\uppi^0$) coherent interactions from the LE beam exposure.
The shaded region shows events with $E_{vis} > 1$\,GeV.
The dashed histogram shows $\etapi$ for incoherent NC production 
of resonances that decay into single-$\uppi^0$ channels.  The latter distribution is shown area-normalized
to the signal distribution to elicit differences in shape.}
\label{Fig04}
\end{figure}

Monte Carlo distributions without selections are shown in Figs.~\ref{Fig03} and~\ref{Fig04}
for kinematic variables of Reaction \eqref{eq:nccoh_reaction}.
The shaded portions of these distributions denote events that have $E_{vis}$ greater than 1.0\,GeV.  
The remaining events (clear histogram regions) 
cannot be reliably identified as EM shower events and are excluded from the analysis.
The distribution of  $\cos\theta_{shw}$ for coherent events (Fig.~\ref{Fig03}a)
is sharply peaked, with 61\% of the total sample having $\cos\theta_{shw} > 0.97$.
The distribution of visible energy, $E_{vis}$, peaks below 1.0\,GeV and falls with increasing 
energy (Fig.~\ref{Fig03}b).   It is predicted that 48\% of signal events deposit more than 1.0\,GeV, and that
93\% have $E_{vis}$ less than 4.0\,GeV.  
The $\cos\theta_{shw}$ and $E_{vis}$ distributions
reflect a peaking of signal events at low values of $\etapi$, as is apparent in
Fig.~\ref{Fig04}, where NC($\uppi^0$) coherent events  are clustered at 
$\etapi \leq 0.050$\,GeV.  Broader $\etapi$ distributions are
predicted for incoherent NC reactions with topologies dominated by EM showers.

\section{Background Reactions}
\label{sec:BgrNeuReactions}

Background interactions originate from one of four
neutrino reaction categories, namely NC, $\CCnm$, $\CCne$, and purely leptonic interactions.
It is useful to divide the NC and $\CCnm$ categories according to the final-state hadronic processes 
used in MC modeling.   The relevant processes are resonance production (RES)
and deep inelastic scattering (DIS).   Electromagnetic showering particles 
dominate the reconstructed shower in all of the background categories.

\smallskip
\noindent
{\it Neutral-current reactions.} The dominant background 
arises from non-coherent NC events with final-state neutral pions that deposit significant shower energy 
and little additional energy above the MINOS detection thresholds.    
Their final-state shower angles with respect to the beam, however, are more broadly distributed
than those of NC($\uppi^0$) coherent scattering.
 
\smallskip
\noindent
{\it \CCnm~reactions.}  There is a subset of \CCnm~events in which the muon track is not identified,
and the hadronic shower is dominated by a single $\uppi^0$.

\smallskip
\noindent
{\it \CCne~reactions.}  Beam $\nue$\,($\bar{\nu}_{e}$) neutrinos can initiate
events having single, prompt electrons (positrons) with no evidence
 of recoil nucleons or other hadronic activity.    This $\CCne$ background
is mainly composed of quasi-elastic (QE) scattering, however resonance production and
DIS processes also contribute.  The reconstructed energy distribution
peaks at $\sim$2.0\,GeV, and extends more broadly to higher energies than the distribution of signal events. 
Evidence that the MC simulation accurately describes $\CCne$ quasielastic-like events is provided by the differential 
cross-section measurements of Ref.~\cite{ref:J-Wolcott-minerva}.
 
\smallskip
\noindent
{\it Purely leptonic interactions.}  
A small background arises from purely leptonic interactions that
initiate energetic single electrons or positrons.    
It consists of $\nu_{\mu}$-electron scattering,
together with much smaller contributions from $\nu_{e}$-electron scattering and 
from the corresponding antineutrino-electron reactions. 
These reactions were not included in the NEUGEN3 event generator, 
and so the neutrino generator GENIE~\cite{ref:GENIE} was used as input to a full simulation.   
(A check on this GENIE prediction for the NuMI LE beam 
is provided by a recent MINER$\nu$A measurement~\cite{ref:J-Park-minerva}.)
Purely leptonic scattering is estimated to be 
$1.2\%$ of the selected data sample, and $(9.7\pm 0.8)\%$ of the extracted coherent signal.  
The background amount, calculated for the data POT exposure, 
was subtracted from the $\cos\theta_{shw}$-vs-$E_{vis}$ template of the data prior to further analysis steps.

\section{Event Selection}
\label{sec:EvtSel}

A preselection was applied to the data.   Events were required to have been recorded 
when both detector and beamline were fully operational.    
The shower vertex and cluster of hits were required to be fully contained within the fiducial volume and to have
visible energy above 1 GeV.   Events with multiple showers, multiple tracks, or single tracks longer than 2\,m were rejected.
For events that passed the data quality and the fiducial volume containment requirements, 
the subsequent cut on visible energy removes an estimated 47\% of coherent NC($\uppi^0$) events.   
Additional losses of signal are incurred by removal of events having multiple reconstructed showers 
or having muon-like topologies;  the losses are at the sub-percent level for each of the latter cuts.
A cutoff was imposed on $E_{vis}$ at 8.0\,GeV as few coherent events are predicted to occur above
this value.   Additionally, $K$-decay rather than muon decay begins to dominate $\CCne$ production above 8.0\,GeV.
This means that regions above 8.0\,GeV are not predictive of background in the signal region and cannot be used as sidebands.
This requirement is estimated to remove 2.9\% from the total signal (including signal with $E_{vis} < 1.0$\,GeV).

\subsection{Multivariate algorithm classification}
\label{subsec:Multi-alg}
 
Further isolation of candidate events was achieved using a
Support Vector Machine (SVM) classification algorithm. 
The output of the SVM is a discriminant value assigned to each event,
hereafter referred to as the \emph{Signal Selection Parameter} (SSP).  
The SVM output for a set of input variables, or ``attributes",
was developed from training samples of MC events~\cite{ref:Cherdack-Thesis}.
The SVM can accommodate large numbers of input variables whose information
content carries various degrees of redundancy;  its performance improves in
accordance with the total amount of discriminatory information provided.
For this analysis, thirty-one different reconstructed quantities were fed to the SVM for each event.
The variables represented five categories of information:  
shower size,  shower shape, shower fit, 
hadronic activity, and track fit. 
Intentionally omitted were reconstructions of shower direction and shower visible energy.
These observables were reserved for use in the fitting of backgrounds to the data.

The SVM algorithm constructs a border surface in the high-dimensional attribute space.   
The SSP is a measure of ``distance" to the border.   
Signal-like regions and background-like regions receive positive
and negative values respectively;  locations on the border have a value of zero.
Events with energy depositions that have
shower-like clusters, are devoid of vertex activity, and have very few remote hits, are to be found in
locations having positive and larger SSP values.

\begin{figure}
\begin{adjustwidth}{-1.0in}{-1.0in}
\centering
\scalebox{0.42}{\includegraphics[trim = 0mm 1mm 0mm 5mm, clip]{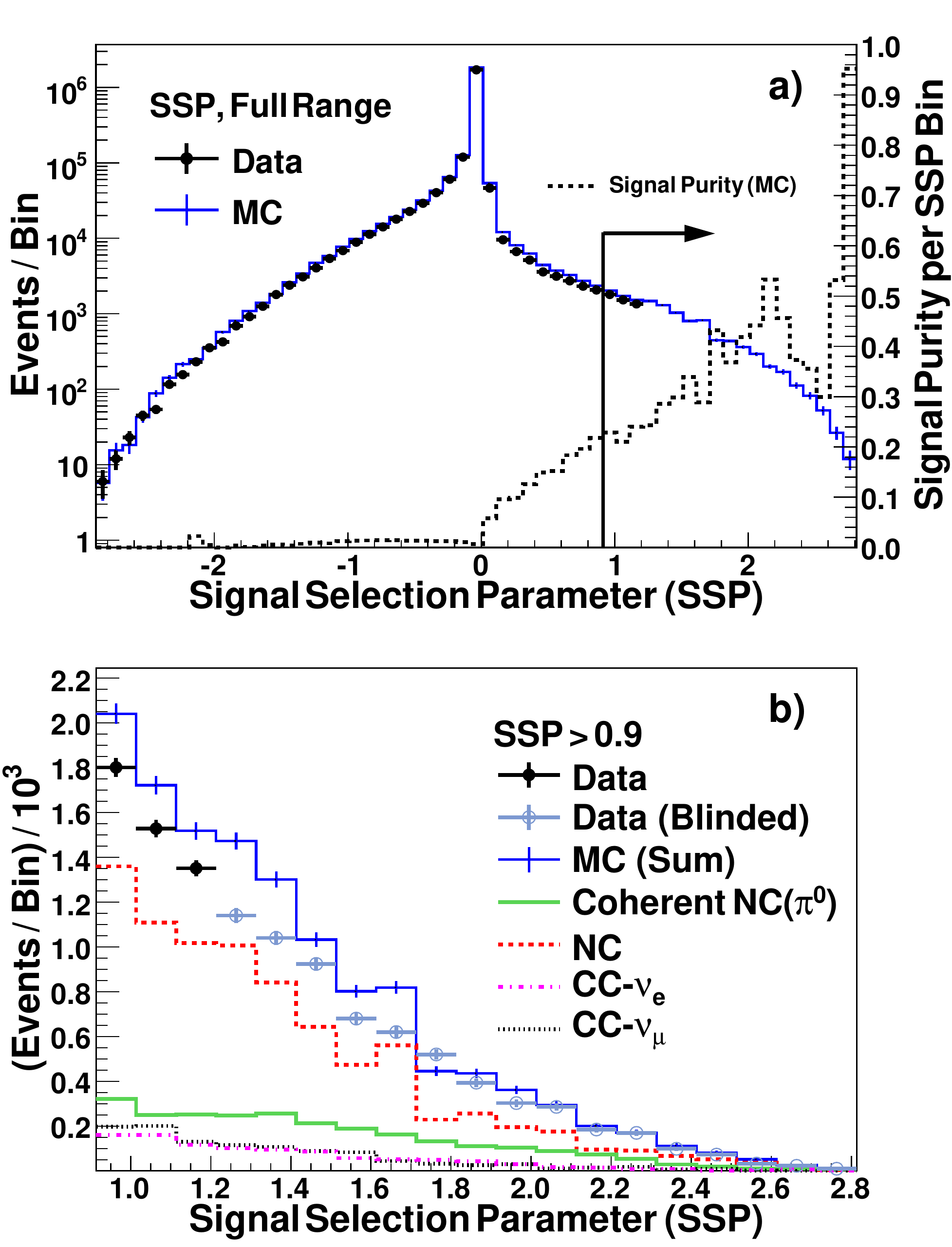}}
\end{adjustwidth}
\caption{(a) Comparison of SSP distribution (left scale) of the unblinded data (black circles) to the MC prediction
(histogram).   The unblinded data were restricted to the region with estimated sigal purity $< 20\%$;
the MC simulation is shown over the full SSP range.
The dashed histogram shows the signal fraction per SSP bin (right scale).
(b) SSP distributions in an interval of enhanced signal content (denoted by the arrow in (a)).   
Histograms show the predicted rate broken out into signal and background contributions.
In the unblinded portion of the signal-enriched region populated by the black-circle data points, 
the simulation reproduces the shape but overestimates the rate of the data.}
\label{Fig05}
\end{figure}

Figure~\ref{Fig05}a compares the SSP distribution of the reference MC sample 
(histogram)  to the unblinded portion of the data (black circles);  display of the latter
distribution is restricted to SSP $<$ 1.2.  The MC signal fraction, or purity, for selected (${\it sel}$) events, 
$\rho$ = $N^{Coh}_{sel}$/ ($N^{Coh}_{sel}+N^{Bkg}_{sel}$), 
is displayed as a function of SSP by the dashed line (with scale to the right).
Figure~\ref{Fig05}b shows the SSP region that is enriched with isolated shower events (SSP $>$ 0.9), 
with the MC simulation broken out into signal and background contributions.
For the region in Fig.~\ref{Fig05}a in the vicinity of SSP = 0 
that contains the bulk of the unblinded data, the simulation matches the data
to within $ 5\%$.   However Fig.~\ref{Fig05}b shows that, 
for the unblinded SSP bins that lie adjacent to the signal-enriched region and contain the black-circle data points,
the MC simulation reproduces the slope of the data but predicts a higher event rate.   
This discrepancy motivates the development of further analysis methods
to constrain the background model using data measurements.  The data in Fig.~\ref{Fig05}b displayed with blue-shade circles
are shown for completeness; their bins were blinded in the analysis.

\subsection{Signal-enriched sample and sidebands}
\label{subsec:Samples}

The background estimation can be significantly constrained using information available 
in sideband samples that lie close to the signal phase-space but have low signal content.
To this end, selections are used to isolate a signal-enriched sample and to define two
separate sideband samples.   These selections are made in two stages.   
In the first stage, a piece-wise linear boundary is defined over the plane 
of SSP versus $\etapi$~\cite{ref:Cherdack-Thesis}.   The boundary defines regions in such
a way as to isolate samples enriched with certain desired properties.   (The specifics
of boundary placement are stated below.)    Two such regions are defined, one contains
the {\it selected sample}, and the other contains the {\it near-SSP sample}.  In the second
stage, the events of the two samples are re-binned as a function of $\cos\theta_{shw}$-vs-$E_{vis}$
and are then separated into regions of high purity (the signal region) and of low purity (the sideband).  
The samples and the selection criteria are elaborated below.

\smallskip
\noindent
{\it The selected sample:}  Events are chosen that populate 
a contiguous region of the SSP-vs-$\etapi$ plane having highest purity and 
containing $\ge 10\%$ of estimated coherent signal events.   
These events (approximately $0.24\%$ of the MC sample shown in Fig.~\ref{Fig05}a)
comprise the selected sample.  Specifically, events of the selected sample 
are required to have SSP $>\,0.5$ for $\etapi\,< 0.2$,
or else SSP $> \text{max}\{(1.3\,-\,4\times\etapi),\,-\,0.9\}$ for $\etapi > 0.2$.  
(An illustrative plot is available as Fig. 6.2 of Ref.~\cite{ref:Cherdack-Thesis}.)

\begin{figure}
\begin{adjustwidth}{-1.0in}{-1.0in}
\centering
\scalebox{0.45}{\includegraphics[angle=0, trim = 0mm 0mm 0mm 14mm, clip]{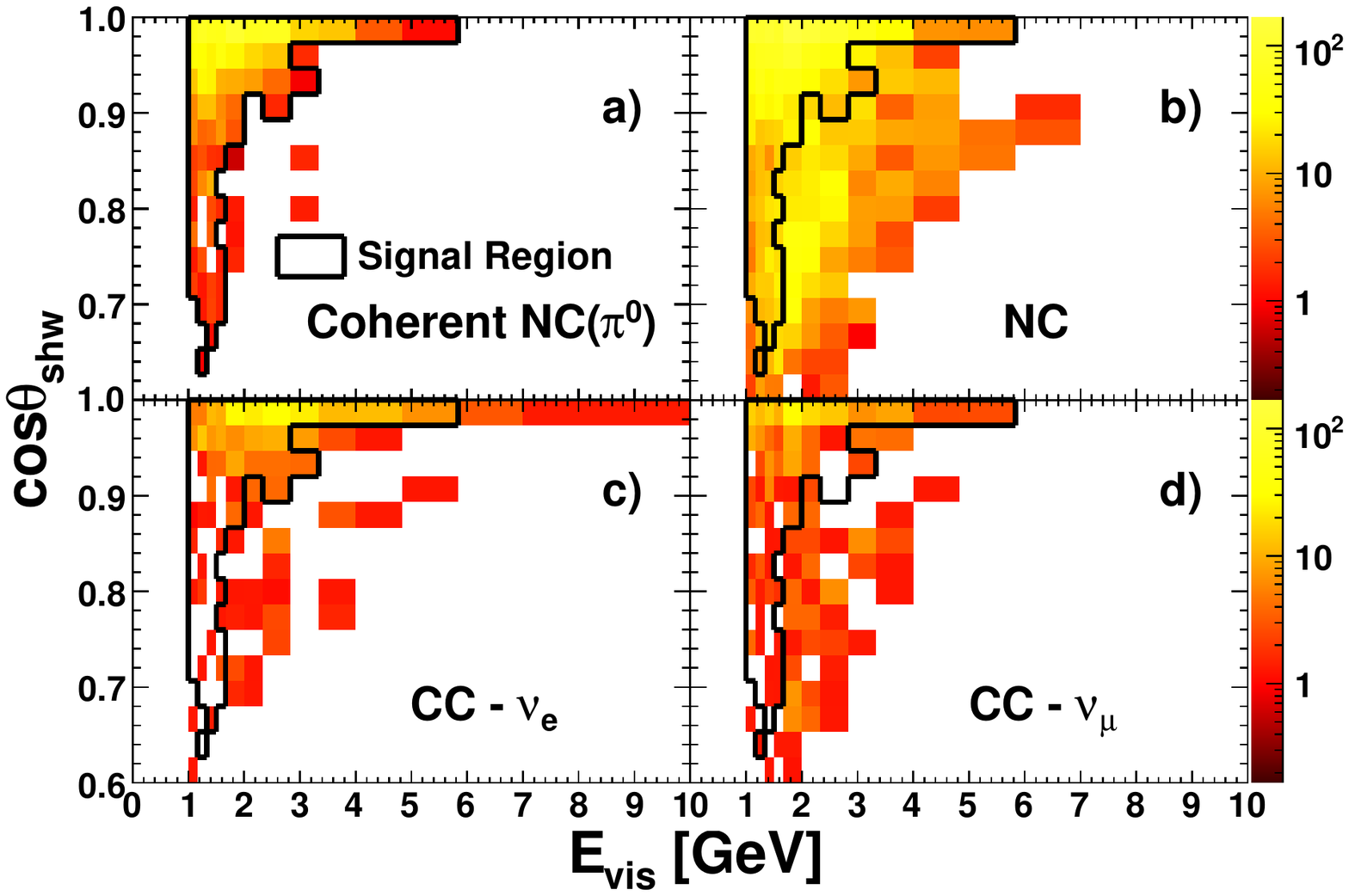}}
\end{adjustwidth}
\caption{Distributions of the MC selected sample 
over the $\cos\theta_{shw}$-vs-$E_{vis}$ plane for (a) signal events, 
(b) the sum of NC resonance plus NC DIS
background templates, (c) the $\CCne$ background template, and (d) the sum of
$\CCnm$ resonance and $\CCnm$ DIS background templates.
Bin-by-bin shading (scale on the right) depicts the event populations.
The signal region enclosed by the solid-line border 
shown on all plots is excluded from fitting to the background.}
\label{Fig06}
\end{figure}

Distributions of the selected MC sample over the $\cos\theta_{shw}$-vs-$E_{vis}$ plane, 
shown separately for signal and backgrounds, are plotted in Fig.~\ref{Fig06}.    
The black-line border separates the bins into two regions according
to their signal purity as described below.   
A large fraction of the sample consists of background events; the relatively large contribution
from NC background can be seen in Fig.~\ref{Fig06}b. 

\begin{itemize}
\item{ {\it Selected sample, signal region:}} The region of the $\cos\theta_{shw}$-vs-$E_{vis}$ plane with bins 
predicted to have $\rho > 5\%$, comprises the signal region of the analysis.    Its outer boundary is shown by
the black-line border superposed on the $\cos\theta_{shw}$-vs-$E_{vis}$ distributions of Fig.~\ref{Fig06}. 

\item{ {\it Selected sample, sideband:}} The selected-sample population lying outside of the signal region on the 
$\cos\theta_{shw}$-vs-$E_{vis}$ plane is predicted to have bins with $\rho < 5\%$.    These events provide information
concerning signal-like backgrounds; they comprise the sideband portion of the selected sample.
\end{itemize}

\noindent
{\it The near-SSP sample:} A second sample, designated the near-SSP sample, 
populates regions adjacent to, but on the opposite side of, the border previously specified that encloses
the selected sample on the SSP-vs-$\etapi$ plane.   Like the selected sample sideband, 
the near-SSP also contains signal-like background events.  
Its inclusion provides additional statistical power to the background fits.

\begin{itemize}
\item{{\it Near-SSP, sideband:}}   There is a region of the near-SSP $\cos\theta_{shw}$-vs-$E_{vis}$ plane 
where the binned event populations have $\rho < 5\%$ in each bin.   
The events that are contained in this region comprise
the near-SSP sideband sample.

\item{{\it Near-SSP, excluded region:}}  The remainder of the near-SSP has purity above 5\% and is excluded from the near-SSP sideband.  
The purity in this region is too low for use as a signal region, as uncertainties 
on the subtracted backgrounds overwhelm the modest gains from statistics.   
Consequently this subsample is excluded from the analysis altogether.
\end{itemize}

\smallskip
As part of the blinding protocol,  data in the two sideband samples 
were not investigated until the sideband fitting procedure 
was fully developed based on mock data studies.   
Similarly, the data in the signal region were not evaluated 
until the fit to the sideband samples was complete, 
and the background rates in the signal region 
and their associated uncertainties were fully determined.

As elaborated in Sec. VII, the templates comprising 
the background model are tuned via fitting to match the data of the sideband samples.   
The background estimate to be subtracted from the data
is thereby anchored in the sidebands but it also encompasses the signal region.
The number of data events in the signal region that exceed 
the estimated background population, represents the coherent scattering signal.

\begin{figure}
\begin{adjustwidth}{-1.0in}{-1.0in}
\centering
\scalebox{0.45}{\includegraphics[trim = 0mm 1mm 0mm 2mm, ]{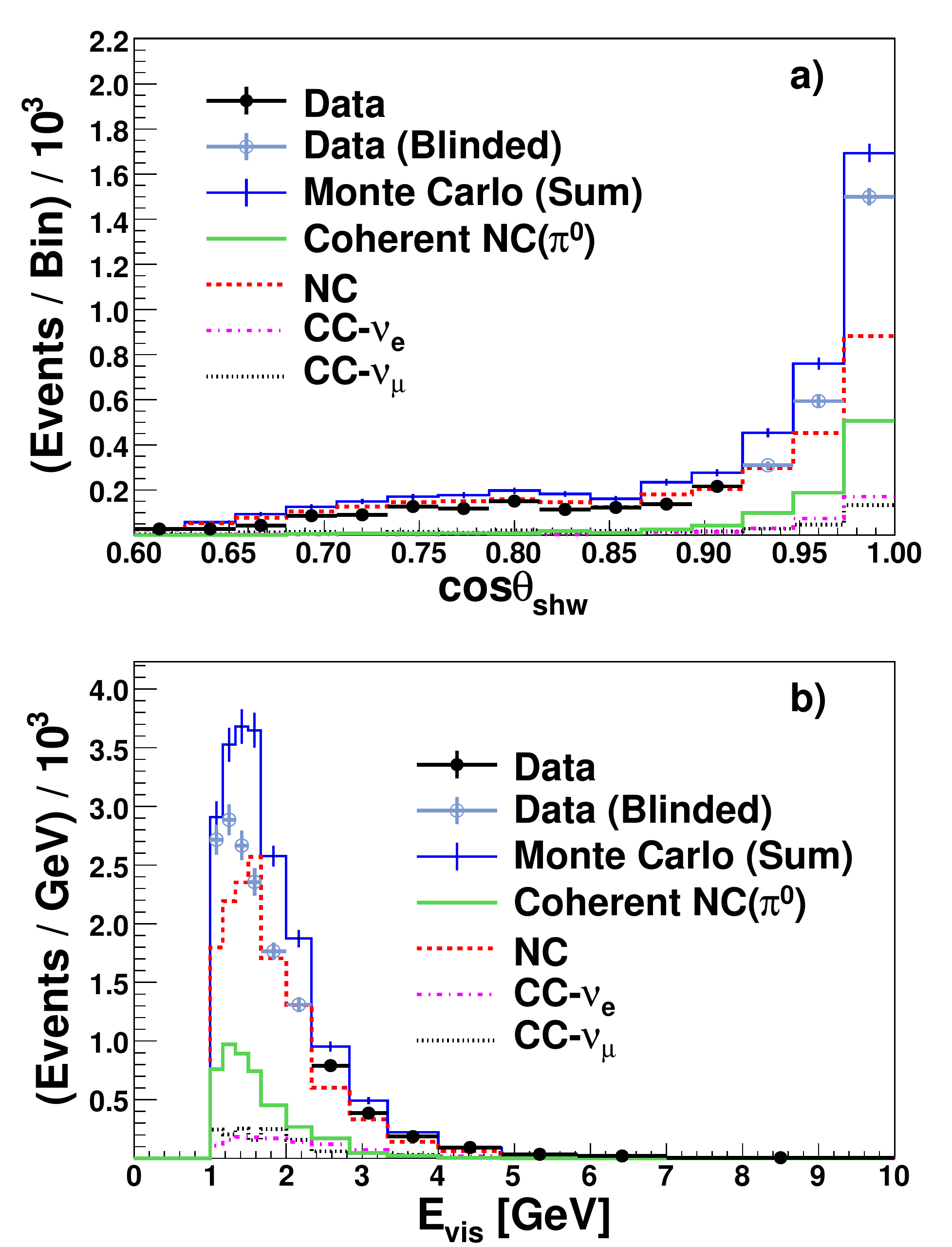}}
\end{adjustwidth}
\caption{Distributions of reconstructed 
 (a) $\cos\theta_{shw}$  and (b) $E_{vis}$ for the data (circles, statistical errors) and 
MC (histograms) of the selected sample.
Data shown as black circles are in the unblinded regions $(\rho < 20\%)$.   The reference MC
matches the shape but exceeds the rate of the unblinded data.}
\label{Fig07}
\end{figure}

Figure~\ref{Fig07} shows the distributions of $\cos\theta_{shw}$ and $E_{vis}$ for the MC selected sample,  
normalized to the data exposure.   These depict projections 
of the distributions shown in Fig.~\ref{Fig06}.
The sample contains 935 coherent NC($\uppi^0$) events
(19.1\% of the sample), together with 3,960 background events. 
The composition of the background is 81.8\% NC, 9.3\% $\CCnm$, and 8.9\% $\CCne$.

\begin{figure}
 \begin{adjustwidth}{-1.0in}{-1.0in}
 \centering
 \scalebox{0.43}{\includegraphics[trim = 0mm 0mm 0mm 0mm]{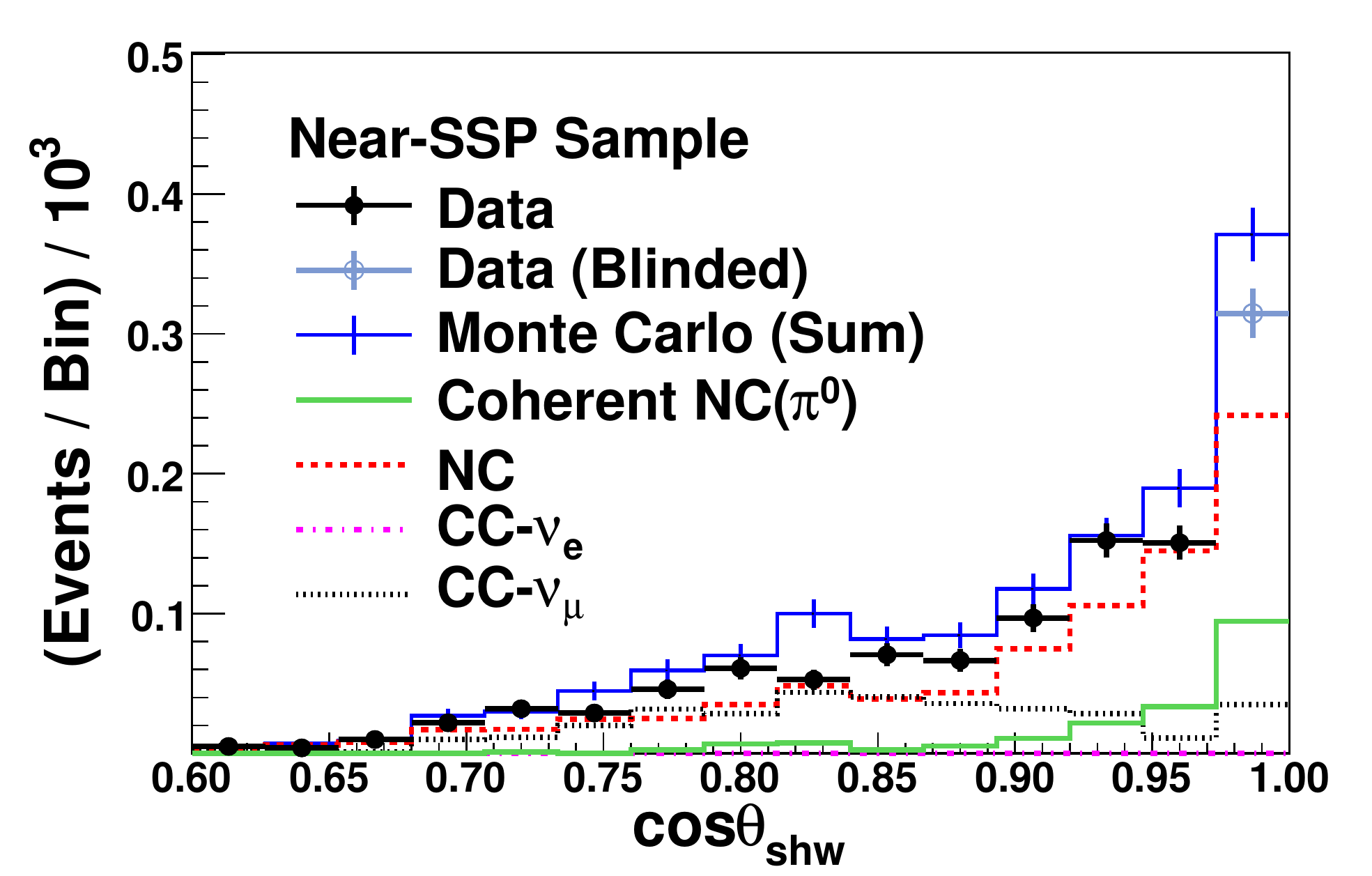}}
 \end{adjustwidth}
\caption{Distribution of reconstructed $\cos\theta_{shw}$ for data (circles) and the reference MC (histograms) of the near-SSP sample.
In the background dominated regions at larger angles, the MC predicts the shape but overestimates the rate of the data.}
 \label{Fig08}
\end{figure}

Figure~\ref{Fig08} shows the $\cos\theta_{shw}$ distribution of the near-SSP sample including the sideband and the excluded region.  
Compared to the selected-sample sideband, the near-SSP sideband has lower event statistics, however its 
lower purity allows a larger number of $\cos\theta_{shw}$-vs-$E_{vis}$ bins to be included. 
Roughly speaking, the near-SSP sideband includes bins with $\eta_{\pi}\geq 0.1$ while the selected-sample sideband
restricts to bins that satisfy $\eta_{\pi}\geq 0.2$.
In both Fig.~\ref{Fig07} and Fig.~\ref{Fig08},  the data shown by the solid circles 
are in the unblinded regions, while the data displayed
as blue-shade circles were blinded.    For the unblinded data,
the MC simulation is seen to overestimate the rate of selected data events
by $\sim 35\%$.    It will be shown that this discrepancy is removed
by adjusting the background models, within uncertainties, to 
match data rates observed in the sideband samples.

\section{Background Estimation by Fitting to Data}
\label{sec:BgrByFitting}

Central to the analysis is its background fitting procedure
 which delivers an effective accounting of most of the systematic 
uncertainties of the measurement using relatively few parameters.    Sections~\ref{sec:BgrByFitting}
through \ref{sec:FitSideData} describe its design and performance.

\subsection{Fit normalization parameters}
\label{sec:fitparams}

For each background category, two separate MC templates containing either selected or near-SSP events
are constructed as two-dimensional $\cos\theta_{shw}$-vs-$E_{vis}$ histograms. 
The bin sizes are set according to experimental resolutions.   
Bins of $E_{vis}$ are proportional to its residual,
$|\,E_{vis} - E_{true}|$, and enlarge with energy to match 
the resolution dependence (residual/$E_{true}\sim$20\%).  
For $\cos\theta_{shw}$, its residual over the sample is nearly
constant and so a constant bin-width of 0.04 is used.
The MC templates, together with similar histograms of the data,
are the principal inputs to the fit.

Figure~\ref{Fig06} shows the MC $\cos\theta_{shw}$-vs-$E_{vis}$ distributions
of selected-sample events for the signal (Fig.~\ref{Fig06}a) and
for the background reaction categories: NC (Fig.~\ref{Fig06}b), $\CCne$ (Fig.~\ref{Fig06}c), and
$\CCnm$ (Fig.~\ref{Fig06}d).  Events enclosed by the solid-line border lie in the signal region while events lying 
outside belong to the sideband.    As previously noted, the NC and $\CCnm$ categories are further divided by the analysis
into sub-categories that distinguish baryon resonance production and DIS interactions.  

Associated with each background reaction category there
is a \emph{normalization parameter};  it serves to scale the total number of events 
assigned to the template distribution of the background.
Studies of fitting using simulated data experiments showed 
the normalization parameters for the templates 
of $\CCnm$ resonance production and NC resonance production 
to be highly correlated.   Strong correlations were also observed 
for the $\CCnm$ DIS and NC DIS templates.
Thus it was decided to combine each of these pairs 
of background categories, allowing for each pair a single template scaled by a 
normalization parameter.    Three templates with independent normalizations 
then suffice to describe the backgrounds: 
\emph{i)} NC and $\CCnm$ resonance production events;
 \emph{ii)} NC and $\CCnm$ DIS events;  and
 \emph{iii)} $\CCne$ events.
Hereafter, the corresponding normalization parameters 
are designated using $n_{res}$, $n_{dis}$, and $n_{\nue}$.

\smallskip
If a systematic error causes the template
normalizations to change, but not the shapes of the 
$\cos\theta_{shw}$-vs-$E_{vis}$ distributions, 
then that error can be absorbed into the 
normalization parameters.  It was demonstrated using
simulated experiments (see Sec.\,\ref{subsec:Evalindividual}) that most 
sources of systematic uncertainty can be accounted for in this way.  
This approach simplifies the treatment and promotes the identification of a 
minimal set of effective systematics parameters.
  
There are two systematic uncertainty sources that can significantly
alter the shapes of the template distributions, namely the energy scale for EM showers and
the assignment of the Feynman scaling variable ($x_F$) to final-state nucleons.  
These sources must be fit for independently, 
and each requires a \emph{systematic parameter} (Sec.\,\ref{subsec:Evalindividual}).

\subsection{Limiting the signal content of sidebands}
\label{sec:SigSide}

It is observed with simulated experiments that signal events in sideband samples
bias the determination of the number of coherent NC($\uppi^0$) events toward the MC prediction.   
It is important to minimize this influence by defining the sidebands such that only bins
with low signal purity are included.   On the other hand, limiting the number of bins in the sidebands
reduces the amount of information available to the fit.  
As a compromise,  the estimated signal purity of bins 
in the sidebands was required to be less than 5\%.
With the latter requirement, this bias, inherent to the analysis fitting procedure, is a  
small effect of 5.8\%.    Its contribution to the signal rate is corrected for, 
and the uncertainty arising from the correction is propagated to the error budget.

\subsection{The $\chi^2$ fit to the background}
\label{ssc:chisq}

Best-fit values for the background normalization parameters $n_{res}$, $n_{dis}$, and $n_{\nue}$ 
plus two systematic parameters (Sec.\,\ref{subsec:Evalindividual}) 
that allow for shape distortions of the background templates,
are determined by minimization of the $\chi^2$:
\begin{equation}
 \begin{split}
   \chi^2 = 2\displaystyle\sum_{i} \left[\left(\ln\displaystyle\frac{N^{Data}_{i}}{N^{MC}_{i}}-1\right)N^{Data}_{i} + N^{MC}_{i}
   \right. \hspace*{.5in}\\
   +
  \left. \left(\ln\displaystyle\frac{N^{MC}_{i}}{N^{adj}_{i}}-1\right)N^{MC}_{i} + N^{adj}_{i}\right]  + ~ \text{penalty} ~\bf{ .} 
   \raisetag{0.8in}
 \end{split}
 \label{eq:chi2}
\end{equation}
\noindent
The $\chi^2$ summation is taken over the bins, $i$, of the selected and near-SSP  
sideband regions of the $\cos\theta_{shw}$-vs-$E_{vis}$ plane. 
The first two terms within the brackets of Eq.\,(\ref{eq:chi2})
represent the likelihood that, according to Poisson statistics, the number of data events of bin $i$ agrees 
with the number of events predicted by the MC simulation. 
Here, $N^{Data}_{i}$ is the number of data events observed in bin $i$, and
$N^{MC}_{i}$ is the number of events expected 
in the same bin for a given set of values for the five parameters of the fit.    

Due to the relatively low rate of coherent NC($\uppi^0$) interactions 
and their associated backgrounds, the selected sample -- although
extracted from very large MC samples -- has limited statistics.
This problem is addressed 
by introducing the third and fourth terms constructed according to the 
method of Beeston and Barlow~\cite{ref:limitMCstats}.   
In brief, the MC content of each bin arising from all the MC samples is fitted to the corresponding data so that the 
sum of terms three and four in \eqref{eq:chi2} is minimized for each bin.   
The logarithmic term imposes a cost
for the adjustment of the MC simulation from $N^{MC}_{i}$ to 
its corresponding fitted value, $N^{adj}_{i}$.  
The inclusion of the latter terms 
effectively replaces $N^{MC}_{i}$ with $N^{adj}_{i}$ plus the penalty.

An additional penalty term in the $\chi^2$ constrains 
the values of the fit parameters;  the constraints are based upon the
studies of systematic uncertainties
discussed in Sec.\,\ref{subsec:Evalindividual}.   
The penalty term is constructed using a covariance matrix which
encodes the variations allowed to the vector of fit parameters, $\hat{\delta}$,    
as related in Sec.\,\ref{sec:Penalty}:
\begin{equation}
\label{eq:penalty_full}
{\mathrm{penalty}} = \vec{\delta} \cdot (V)^{-1} \cdot (\vec{\delta})^{T} \bf{ .}
\end{equation}
\noindent 
Multiple covariance matrices were formulated to allow for asymmetries in the parameter errors.   The appropriate
matrix is chosen based on the sign of the normalization parameter deviations.

\section{Extraction of the Signal Rate}
\label{sec:ExtractSig}

Minimization of the $\chi^2$ yields the best-fit values for the fit parameters, and these are
used to estimate the rate for each category of background events across the entire selected sample.
\subsection{Raw signal event rate}
\label{subsec:Rawsigrate}
The number of selected events
(in bin $i$) contributed by background template $b$ is $N_{ib}^{MC}$.  Each
 $N_{ib}^{MC}$ is scaled by a background normalization parameter, $f_b$ = $n_{res}$, $n_{dis}$, or $n_{\nue}$, 
and the systematic scale factor, $s_{ib}$.    The value of $s_{ib}$ is the sum of fractional changes
(bin-by-bin) induced by changes in value for 
systematic parameters associated with uncertainties of EM energy scale and 
of $x_{F}$ assignment to final-state nucleons (see Sec.\,\ref{subsec:Evalindividual}). 
The predicted number of background events in each bin
$N^{Bkg}_{i}$, is the sum of the scaled values 
of $N_{ib}^{MC}$ over the three background templates:
 
\begin{eqnarray}
N^{Bkg}_{i} = \displaystyle\sum_{b} f_{b}\, s_{ib}\, N_{ib}^{MC} \bf{ .}
\label{eq:sigExtract_bkgs}
\end{eqnarray}

\noindent
The measured signal in each bin, $N^{Coh}_{i}$, is the difference between the number of 
 data events and the number of neutrino background events as estimated using Eq.\,(\ref{eq:sigExtract_bkgs}):
\begin{equation}
 N^{Coh}_{i} = N^{Data}_{i} - N^{Bkg}_{i} \bf{ .}
\label{eq:sigExtract_sig}
\end{equation}
The background subtraction yields a count of measured signal events 
in each bin of the selected sample.

\subsection{Acceptance corrections}
\label{subsec:AcceptCorrect}

The acceptance correction is applied via an efficiency function,
\begin{eqnarray}
\epsilon_{i} = \frac{N^{MCs}_{i}}{N^{MCt}_{i}} \textbf{ ,}
\label{eq:sigExtract_eff}
\end{eqnarray}
where $N^{MCs}_{i}$ is the number of coherent 
NC($\uppi^0$) MC events in bin $i$ in the selected 
sample and $N^{MCt}_{i}$ is the total number of 
coherent NC($\uppi^0$) events in bin $i$ predicted by the reference MC.

There are a small number of bins for which very few 
signal events are estimated and the efficiency approaches zero.
These bins are omitted from the sum-over-$i$ and their correction is applied via an 
overall factor $\epsilon_0^{-1}$, calculated as the ratio 
of the (predicted) total signal rate divided by the selected signal rate for all bins with non-zero efficiency.
The choice of an acceptance correction for each bin, 
either bin-by-bin ($\epsilon_{i}$) or overall ($\epsilon_0$), was determined by minimizing
the uncertainty propagated to the measured signal.   Also included in $\epsilon_0$
is the correction for signal loss incurred by the $E_{vis} < 8.0$ GeV cutoff, 
and a small correction for interactions that were not properly reconstructed.

There are coherent NC($\uppi^0$) MC events with true visible energy below 1.0 GeV 
that reconstruct with $E_{vis}>1.0$\,GeV, and vice versa. 
An additional correction is applied as a weight factor, $\xi$, to account for  the net event migration 
across the cut boundary at $E_{vis} = 1.0$\,GeV.
The acceptance-corrected coherent event rate is then 
\begin{equation}
N^{Coh} = \frac{\xi}{\epsilon_{0}} \displaystyle\sum_{i}^{\epsilon_{i}>0} \frac{1}
{\epsilon_{i}} \Bigg( N^{Data}_{i}  \, - 
\left. \displaystyle\sum_{b} f_{b}\, s_{ib}\,  N_{ib}^{MC} \right) \,\bf{.}
\raisetag{0.4in}
\label{eq:sigExtract_ER}
\end{equation}
\noindent
The integrated effect of the bin-by-bin acceptance corrections $\epsilon_{i}^{-1}$ in the summation of Eq.~\eqref{eq:sigExtract_ER} 
is equivalent to to an overall correction of about 8.2.
The factors $\epsilon_0^{-1}$ and $\xi$ in Eq.\,(\ref{eq:sigExtract_ER}) introduce corrections of 1.42 and 
0.90 respectively;  their net effect is to shift the calculated signal rate upward by $28.1\%$.

\section{Systematic Uncertainties}
\label{sec:SystUnc}

Sources of systematic uncertainties are described below.   
The effects of individual sources are summarized in Sec.\,\ref{subsec:Evalindividual}.   
Many of the sources were studied in 
previous MINOS analyses~\cite{ref:MINOS_nue, ref:POT_Error}.

\subsection{Uncertainties in neutrino-interaction modeling}
\label{sec:SourceUncertain}

\noindent
{\it Modeling of $\nu$N cross sections:}
The dominant uncertainties in the cross section model are associated with
$i$) the axial mass $M^{QE}_A$ used in quasielastic cross sections,
$ii$) the axial mass $M^{Res}_A$ used in resonance production cross sections,
and $iii$) the treatment of the transition region between resonance production 
and DIS~\cite{ref:neugen3, ref:D-Bhat_Thesis}.    The values and $\pm1\upsigma$
uncertainties of the model parameters were taken from 
previous MINOS investigations~\cite{ref:POT_Error, ref:CCinclusive}.
The axial masses $M^{QE}_A$ and $M^{Res}_A$ used with dipole form factors 
are effective parameters whose assigned fractional errors makes allowance for
uncertainties arising from nuclear medium effects neglected by the MC such as
2-particle 2-hole excitations and long-range correlations~\cite{martini-2p2h, valencia-2p2h}.

\smallskip
\noindent
{\it Modeling of hadronization:}
Uncertainties in the NEUGEN3 hadronization model reflect
a lack of data on the DIS channels selected by
the analysis.    Six model parameters were identified as having uncertainties that 
influence the predicted event samples and their effects were individually investigated:
\emph{i)} The assignment of Feynman-$\mathit{x}$ to the final-state baryon, $\mathit{x_{F}}$;
\emph{ii)} the probability for $\uppi^0$ production, $\mathit{P(\piz)}$;
\emph{iii)} the correlation between produced neutral-particle multiplicity
and charged-particle multiplicity, $\mathit{n_{0}}$ and $\mathit{n_{\pm}}$, respectively;
\emph{iv)} differences between generator simulations of hadronic systems, 
\emph{gen-diff},  (GENIE~\cite{ref:GENIE} vs NEUGEN3~\cite{ref:neugen3});
\emph{v)}  damping algorithm for transverse momenta, $\mathit{p_{\small{T}}~damping}$; and
\emph{vi)} neglect of correlations which may arise with two-body decays, $\mathit{decay~param}$.
 
\smallskip
\noindent
{\it Intranuclear rescattering:}

Neutrino-induced pions and nucleons can undergo final-state interactions (FSI)
prior to emerging from the parent nucleus.   The analysis accounts for FSI processes
in all incoherent neutrino scattering interactions using a cascade model to simulate
the propagation of produced hadrons within the target nuclei~\cite{ref:Intranuke-Model}.  
For coherent signal reaction (\ref{eq:nccoh_reaction}) however, the rate and final-state momenta
of produced $\piz$s in simulation are taken directly from the Berger-Sehgal model.  The model
accommodates the attentuation of coherently produced $\piz$s by the parent nucleus
by using pion-nucleus elastic-scattering cross sections as input~\cite{ref:BS}.

 The performance of the FSI cascade model
is governed by two types of adjustable parameters.
The first type establishes relative rates for the 
possible intranuclear processes with $\pm 1\upsigma$ as
evaluated for the MINOS analysis of $\numu$ disappearance~\cite{ref:POT_Error}. 
The seven parameters of the first type are \emph{i)} pion charge exchange,
\emph{ii)} pion elastic scattering,
\emph{iii)} pion inelastic scattering,
\emph{iv)} pion absorption,
\emph{v)} $\uppi$-nucleon scattering yielding two pions,
\emph{vi)} nucleon knockout from the target nucleus, and 
\emph{vii)} nucleon-nucleon scattering with pion production.
Parameters of the second type govern the overall rate of intranuclear rescattering, {\it e.g.}
the pion-nucleon cross section and the formation time, $\mathit{T_{formation}}$, for
directly produced hadrons.    

\subsection{Implications for background reactions}
\label{subsec:Bkgd-mod-incert}
\noindent
{\it NC reactions:}  
Generation of NC events is affected by all of the above-listed modeling uncertainties
~\cite{ref:CalDet-2006, ref:T-Yang_Thesis, ref:Boehm_Thesis}. 

\smallskip
\noindent
{\it $\CCnm$ reactions:}   
The cross sections for $\CCnm$ reactions are better-known 
and hence better-constrained than for NC channels.
Moreover the selected $\CCnm$ rate is only 10\% of the selected NC event rate.
Consequently the effects of uncertainties with modeling of $\CCnm$ interactions are 
sufficiently weak to be subsumed by the error range assigned by the fit to the $n_{res}$ and $n_{dis}$
normalization parameters (see Sec.\,\ref{subsec:Evalindividual}).

\smallskip
\noindent
{\it $\CCne$ reactions:} 
The electron-induced showers of selected CC-$\nu_{e}$ events have no visible hadronic activity, 
and so uncertainties arising from hadronization and intranuclear rescattering are negligible.   
The dominant uncertainty in the CC-$\nu_{e}$ event rate
arises from limited knowledge of the ($\nu_{e} + \bar{\nu}_{e}$) flux 
in the NuMI LE beam~\cite{ref:MINOS_nue, ref:Ochoa_Thesis}.
The additional 20\% flux uncertainty is propagated to the uncertainty assigned to
the $\nu_{e}$ normalization parameter.

\subsection{Uncertainties of energy scale and signal model}
\label{scc:cohscattm}

Uncertainties in the electromagnetic energy scale, $\mathit{E_{scale}^{EM}}$, 
and detector calibration contribute significantly to the error budget.
An overall uncertainty of $\pm$5.6\% is assigned to the EM energy scale, reflecting
uncertainty with hadronic contributions to MINOS shower topologies
($\pm5.1\%$), together with uncertainties in the detector response to EM showers ($\pm$2.0\%).
The latter response was evaluated using measurements 
obtained with the MINOS Calibration Detector~\cite{ref:Boehm_Thesis}.

Inaccuracies in modeling the coherent-scattering signal 
can influence the signal amounts inferred from the background levels established by fitting.
Signal model inaccuracies also enter into the acceptance corrections.
The effect was evaluated using simulated experiments employing alternate models 
of the coherent interaction cross section~\cite{ref:Cherdack-Thesis}.
The definitions of the signal region and sidebands, by design,
minimize the influence of the signal model.  
The net effect to the signal rate is accounted for 
by the uncertainty on the $5.8\%$ sideband biasing correction 
of Sec.\,\ref{sec:SigSide} plus the $\pm$3.2\% uncertainty attributed to the acceptance
corrections of Sec.\,\ref{subsec:AcceptCorrect}.


\subsection{Evaluation of sources}
\label{subsec:Evalindividual}

The effect of each source of systematic uncertainty was evaluated individually.  
Monte Carlo samples were created in which a single input parameter, corresponding
to one of the sources (Sec.\,\ref{sec:SystUnc}\,A,C),
was changed by its $\pm$1$\upsigma$ uncertainty.
The $\cos\theta_{shw}$-vs-$E_{vis}$ distribution of each altered sample
was then compared to the background model.
Fitting the sidebands of the background templates to the
sidebands of the altered MC sample yields a re-expression of the $\pm1\upsigma$
uncertainties on the underlying model parameters as uncertainties on template normalizations.
However, for systematic uncertainties that induce changes in the background templates that cannot be adequately
described by normalization changes, use of their underlying model parameters is retained and their
effects on the normalization parameter uncertainties are not included.

For each of the above-described exercises the altered MC distribution was treated like ``data",
thus the altered MC samples are referred to as \emph{Single-Systematic Mock Data} (SSMD).
The overall campaign was to generate SSMD samples for each systematic, 
subject each sample to the template fit procedure that constrains the background
model in the sidebands, and evaluate the outcomes.
Evaluations external to the fitting are used for systematic uncertainties 
associated with calibration (see Sec.\,\ref{sec:Xsec}).     

More specifically, the steps were as follows:    {\it  i)}  For each source of uncertainty,
fluctuations of $\pm 1\upsigma$ in the corresponding parameter induce changes to
the SSMD event distribution in $\cos\theta_{shw}$-vs-$E_{vis}$;  
{\it  ii)}  the changes in the event distribution are evaluated by fitting
the background model to the fluctuated distribution, 
allowing the three background normalization parameters 
$n_{res}$, $n_{dis}$, and $n_{\nue}$, to float without restriction;
{\it  iii)} the SSMD fit result is used to identify whether or not a source introduces a shape change 
into the $\cos\theta_{shw}$-vs-$E_{vis}$ spectrum.  
{\it iv)}  In the cases where the systematic uncertainty does not induce
a significant shape change (most do not), the best-fit values of the SSMD fit 
are used to calculate the allowed variances on (and covariance between)
the normalization parameters in the final analysis fit to sideband data.

For each source of systematic uncertainty, the shifts $-1\upsigma$, $+1\upsigma$ were considered separately.
The fitting to SSMD samples provided the $\chi^2/ndf$ for the best fit,
the fit values for the normalization parameters,  and
the extracted signal, which was compared to the 
value for the reference MC.  Since each SSMD sample is created by inducing a $1\upsigma$ change
in a single systematic parameter, and does not include any statistical fluctuations,
the $\chi^{2}/ndf$ is rated against 0.0 rather than the usual 1.0.

For fifteen of the twenty-two systematic error sources, the SSMD trials yielded $\chi^2/ndf< 0.05$,
well-understood deviations of background normalizations from their nominal values,
and extracted signal event counts which were within $\pm 19\%$ of the 
simulation ``truth" values.  Thus, in fitting the background model to data, 
shifts of these fifteen sources can be absorbed by the normalization parameters.
Typical of these fifteen ``well-behaved" sources is the axial-vector mass, $\mathit{M_{A}^{Res}}$, 
which is here singled out as an example.  The results from an SSMD trial wherein $\mathit{M_{A}^{Res}}$
was subjected to a +1\,$\sigma$ shift are summarized in the bottom row of Table~\ref{tab:syt_rank}.

\smallskip
\begin{table}
\centering
\scalebox{0.93}{%
\begin{tabular}{|c|c||c|c|c|c|c|c|}
\hline
\multirow{3}{*}{\bf{Systematic}} & \multirow{3}{*}{\bf{Shift}} & \multicolumn{3}{|c|}{\bf{Best-Fit Norm.}}     & \multicolumn{3}{|c|}{\bf{\,Fit Outcome\,}}\\
\cline{3-8}
             &     & \multicolumn{2}{|c|}{\bf{NC\small{+}CC$\nu_\mu$}} & {\bf{CC$\nu_e$}} & \multirow{2}{*}{$\bf{\chi^{2}}/ndf$} & \bf{Signal} \\
\cline{3-5}
  \bf{Source}             &     &~$n_{res}$~       & $n_{dis}$        &      $n_{\nue}$          &                                      & \bf{Ratio}  \\
\hline  \hline
$\mathit{E_{scale}^{EM}}$        & -5.6\%          & 1.13 & 0.75 & 0.85 & 1.32 & 1.45 \\
$\mathit{E_{scale}^{EM}}$        & +5.6\%          & 1.00 & 1.20 & 1.33 & 0.79 & 0.91 \\
$\mathit{x_{F}}$                 & +1 $\upsigma$     & 1.45 & 0.83 & 0.97 & 0.78 & 0.30 \\
$\mathit{n_0(n_{\pm})}$          & $\pm$1 $\upsigma$ & 1.15 & 0.83 & 0.78 & 0.18 & 0.92 \\
$\mathit{decay~param.}$          & +1 $\upsigma$     & 1.15 & 0.85 & 0.90 & 0.15 & 0.78 \\
$\mathit{gen-diff}$             & $\pm$1 $\upsigma$ & 1.03 & 0.95 & 0.80 & 0.13 & 0.87 \\
$\mathit{p_{\small{T}}~damping}$ & $\pm$1 $\upsigma$ & 1.00 & 0.98 & 0.78 & 0.12 & 0.88 \\
$\mathit{T_{formation}}$         & -50\%           & 1.00 & 0.88 & 0.85 & 0.10 & 0.94 \\
$\mathit{T_{formation}}$         & +50\%           & 1.08 & 1.05 & 1.10 & 0.07 & 1.32 \\
$\mathit{M_{A}^{Res}}$          & +15\%           & 1.83 & 1.00 & 1.10 & 0.02 & 1.01 \\ 
\hline  
 \end{tabular}}
\caption{Summary of SSMD studies of uncertainty sources
having potential  to alter the shape of $\cos\theta_{shw}$-vs-$E_{vis}$ distributions.   
Shown are the effects of -1$\upsigma$ and +1$\upsigma$ changes in the sources on the three normalization parameters, 
together with the $\chi^{2}$ and the ratio of the extracted signal to the true signal.  Deviations of the fit
parameters and signal ratio are relative to the nominal value of 1.0 for the reference MC.   The EM energy scale and
$x_F$ of the hadronization model exhibit the most significant effect on the expected number 
of events as a function of $\cos\theta_{shw}$-vs-$E_{vis}$.}
\label{tab:syt_rank}
\end{table}

\subsection{Systematic parameters; fit penalty term}
\label{sec:Penalty}

For the remaining systematic sources shown in Table~\ref{tab:syt_rank}, somewhat larger
$\chi^{2}/ndf$ or excursions of the measured event rates from the reference MC values
were observed.   Table~\ref{tab:syt_rank} lists the SSMD fit results for each of the latter sources of
uncertainty; the sources are ranked according to the reduced $\chi^2$.    In particular,
three of 1$\upsigma$ shifts in two sources have $\chi^{2}/ndf$ which are distinctly worse than the rest.  
The sources  are the EM energy scale (large $\chi^{2}/ndf$ for both -1$\upsigma$ and +1$\upsigma$ shifts), and the parameter associated 
with assignment of $x_F$ to nucleons of final-state hadronic systems (large $\chi^{2}/ndf$ for +1$\upsigma$ shift).  
Their SSMD fit results are displayed in the first three rows of Table~\ref{tab:syt_rank}.

The $\chi^{2}/ndf$ values of Table~\ref{tab:syt_rank} provide guidelines for the introduction of
additional systematic parameters that may entail distortions to template shapes.  
Studies utilizing ensembles of  ``realistic" mock data experiments (see the Appendix)
examined the performance of fit-parameter configurations wherein various combinations of parameters listed
in Table~\ref{tab:syt_rank} were introduced.   The width of the ($N_{fit} - N_{input}$) spectrum obtained from each mock data ensemble
was used as the figure-of-merit for distinguishing among parameter sets.   It was observed that the width was reduced 
with addition of a systematic parameter to account for variation in the EM energy scale, and was further reduced 
when a systematic parameter to account for variation in the
assignment of $\mathit{x_{F}}$ to final-state nucleons was included.
Neither the addition of more parameters nor the utilization of other shape parameter combinations 
yielded a further decrease in the spectral width.

The aggregate of $\pm1\upsigma$ uncertainty from the ensemble of sources of systematic uncertainty 
evaluated by the SSMD trials determines the correlated ranges of variation
to be allowed to the background normalization parameters.   
This greatly reduces the number of systematic parameters that, if otherwise included, would exert
degenerate effects on the predicted background $\cos\theta_{shw}$-vs-$E_{vis}$ distributions.
The resulting fit to data sidebands is less susceptible to multiple minima and less dependent
on the details of the background cross section models.    The above-mentioned ranges are 
enforced in the fit $\chi^2$ of Eq.\,\eqref{eq:chi2} by the penalty term of Eq.\,\eqref{eq:penalty_full}.

\section{Fitting to Data Sidebands}
\label{sec:FitSideData}

\begin{figure}
 \begin{adjustwidth}{-1.0in}{-1.0in}
 \centering
 \scalebox{0.45}{\includegraphics[trim = 0mm 0mm 0mm 10mm]{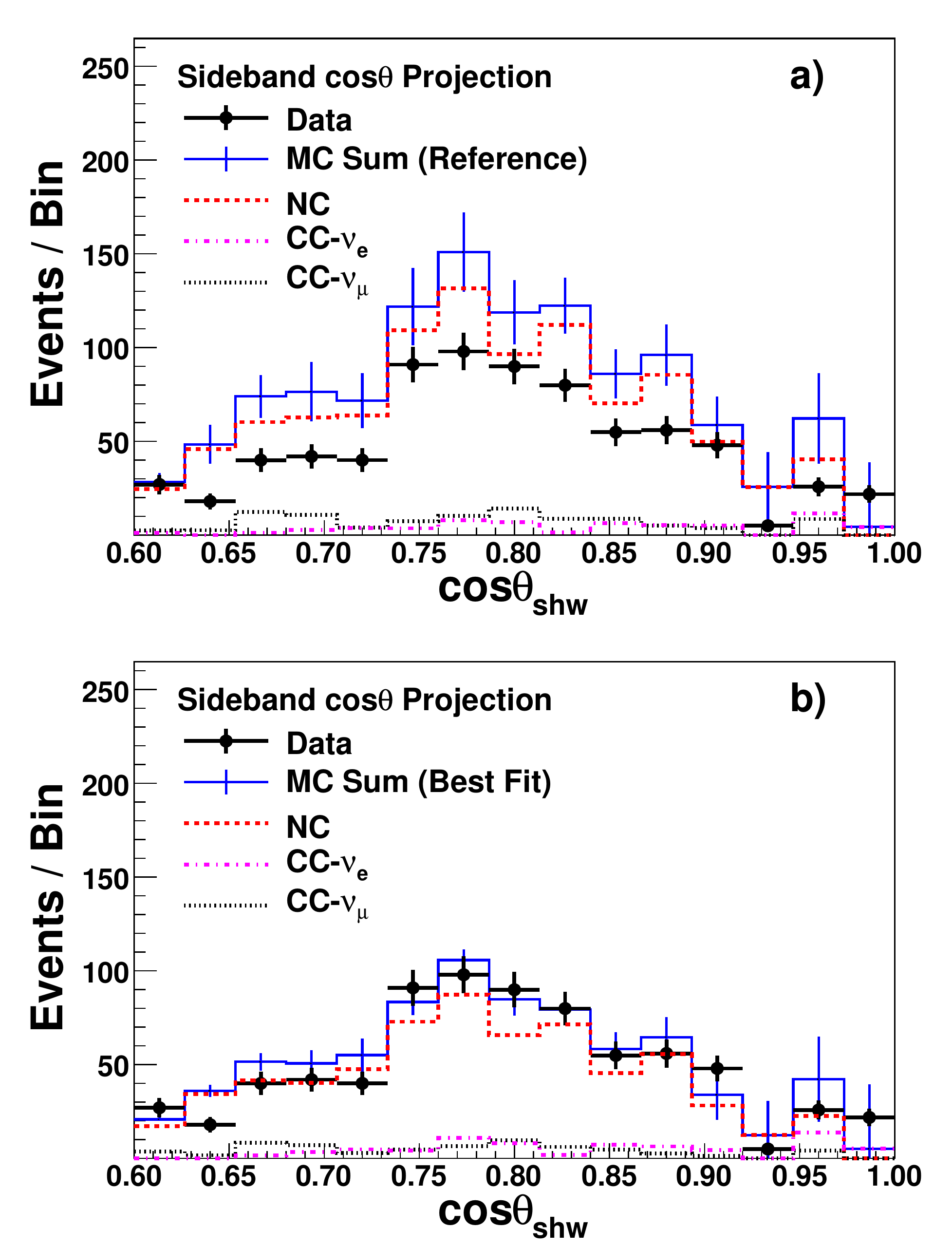}}
 \end{adjustwidth}
 \caption{Distribution in  $\cos\theta_{shw}$ for selected data
 (solid circles) in the sideband region of the
 $\cos\theta_{shw}$-vs-$E_{vis}$ plane. (a) Data versus
 MC background templates prior to fitting.      (b) Data compared to background 
of the best-fit MC.   The fit adjustment
reduced the NC background (dashed) and increased the $\nue$ background 
(dot-dashed)  to achieve a good description (thick-line histogram).}
 \label{Fig09}
\end{figure}

A simultaneous fit over the data of the selected-sample  
and near-SSP sidebands is now carried out 
via minimization of the $\chi^2$ function of Eq.\,\eqref{eq:chi2}.   
The $\chi^2$ uses the three background normalization parameters 
and the two systematic parameters in conjunction with the fit penalty term as described in 
Sec.~\ref{sec:BgrByFitting}.   The fit result establishes
the background prediction in the signal region of the selected sample.
The outcome of the fit is illustrated in Fig.~\ref{Fig09}.
Here, data of the selected sample is compared to the neutrino background model 
for the  $\cos\theta_{shw}$ projection of the sideband region of the $\cos\theta_{shw}$-vs-$E_{vis}$ plane.
The shapes of the distributions in Fig.~\ref{Fig09} reflect
the irregular contour of the sideband region (as indicated by Fig.~\ref{Fig06}).
Figure~\ref{Fig09}a shows the $\cos\theta_{shw}$ projection prior to fitting.  The neutrino NC category
 is the dominant background;  its distribution (dashed line) approximates the shape of the 
sideband data (solid circles), however its normalization is too high
by $\sim35\%$ as noted in Sec.\,\ref{sec:EvtSel}.

Figure~\ref{Fig09}b shows the best fit (solid line histogram)
together with the background composition.    The fit reduced 
the normalizations $n_{res}$ and $n_{dis}$                
by -1.04\,$\upsigma$ and -1.08\,$\upsigma$ respectively (corresponding to 35\% and 25\% reductions),  while increasing
$n_{\nue}$ by +0.40\,$\upsigma$ (a 17.5$\%$ increase).
Additionally the EM energy scale is shifted upwards by +0.15\,$\upsigma$, corresponding 
to a 0.84\% increase in the conversion from energy deposition in the detector to the measured energy in GeV.
The best-fit value for baryonic $x_F$ corresponds to a +0.35$\upsigma$ shift from nominal. 
This change increases the probability that the final-state nucleon 
will emerge in the forward hemisphere of the target rest frame.

The fit to the data gives a reduced $\chi^{2}$ lower than the values obtained in 52.7\% 
of the realistic mock data experiments described in the Appendix,
indicating that the MC simulation is representative of the data to within the MC uncertainties.
Comparisons of the best-fit background to the $E_{vis}$ projection
of the selected data in the sideband, and to the $\cos\theta_{shw}$ and $E_{vis}$ projections 
of the near-SSP data sample, also show a satisfactory description 
of the data~\cite{ref:Cherdack-Thesis, ref:Cherdack-NuInt11}.

\section{Signal Rate and Uncertainty Range}
\label{sec:Xsec}

With the best fit over the data sidebands in hand, the background is set for the
entire $\cos\theta_{shw}$-vs-$E_{vis}$ plane.    At this point  the background prediction 
is fully determined and the data of the signal region is unblinded.

\begin{figure}
 \begin{adjustwidth}{-1.0in}{-1.0in}
 \centering
 \scalebox{0.43}{\includegraphics[trim = 0mm 0mm 0mm 0mm]{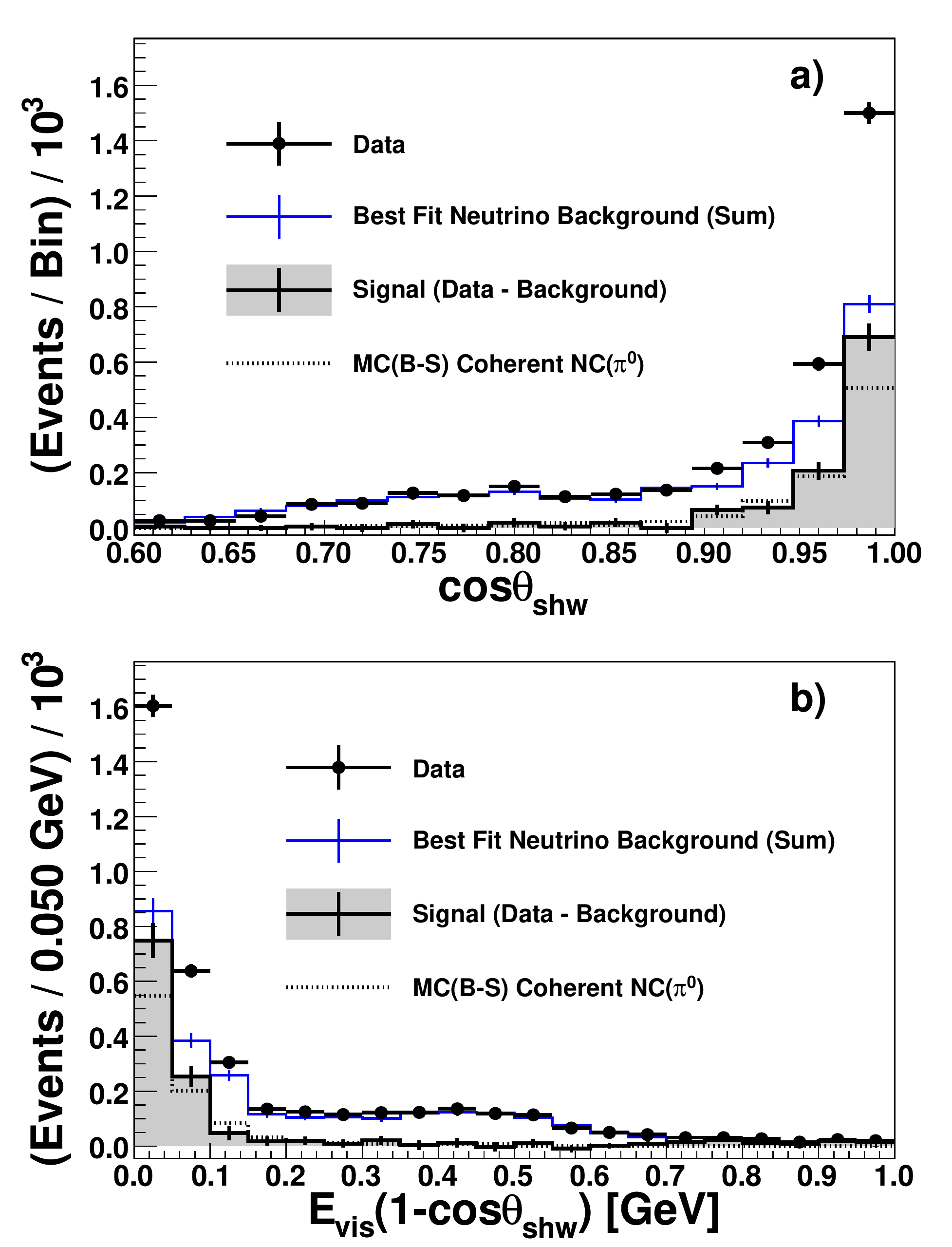}}
 \end{adjustwidth}
\caption{Distributions of candidate NC($\uppi^0$) coherent scattering
events (solid circles, statistical error bars) in $\cos\theta_{shw}$ (a) and $\etapi$ (b). 
The data are compared to the estimate for the neutrino backgrounds (thin blue-line histograms).  
The coherent scattering signal (data minus background) is shown by the shaded histogram,
 together with the signal rate inferred using the Berger-Sehgal model (dotted histogram).}
 \label{Fig10}
\end{figure}

Figure~\ref{Fig10} shows the distributions in $\cos\theta_{shw}$ (Fig.~\ref{Fig10}a) and in
$\etapi$ (Fig.~\ref{Fig10}b) for all selected data.   The predicted background (clear histogram)
shows good agreement with the data points (solid circles) over the lower range ($<\,0.9$) of $\cos\theta_{shw}$
and over the upper range ($> 0.25$) of $\etapi$.   The signal for Reaction\,\eqref{eq:nccoh_reaction} emerges
with either variable as the incident neutrino direction is approached, appearing as
a data-minus-background excess (shaded histograms). 
The errors on the extracted signal are the quadrature sum of errors 
from the background fit plus statistical uncertainties of the data and MC.

\begin{figure}
 \begin{adjustwidth}{-1.0in}{-1.0in}
 \centering
 \scalebox{0.46}{\includegraphics[angle=0]{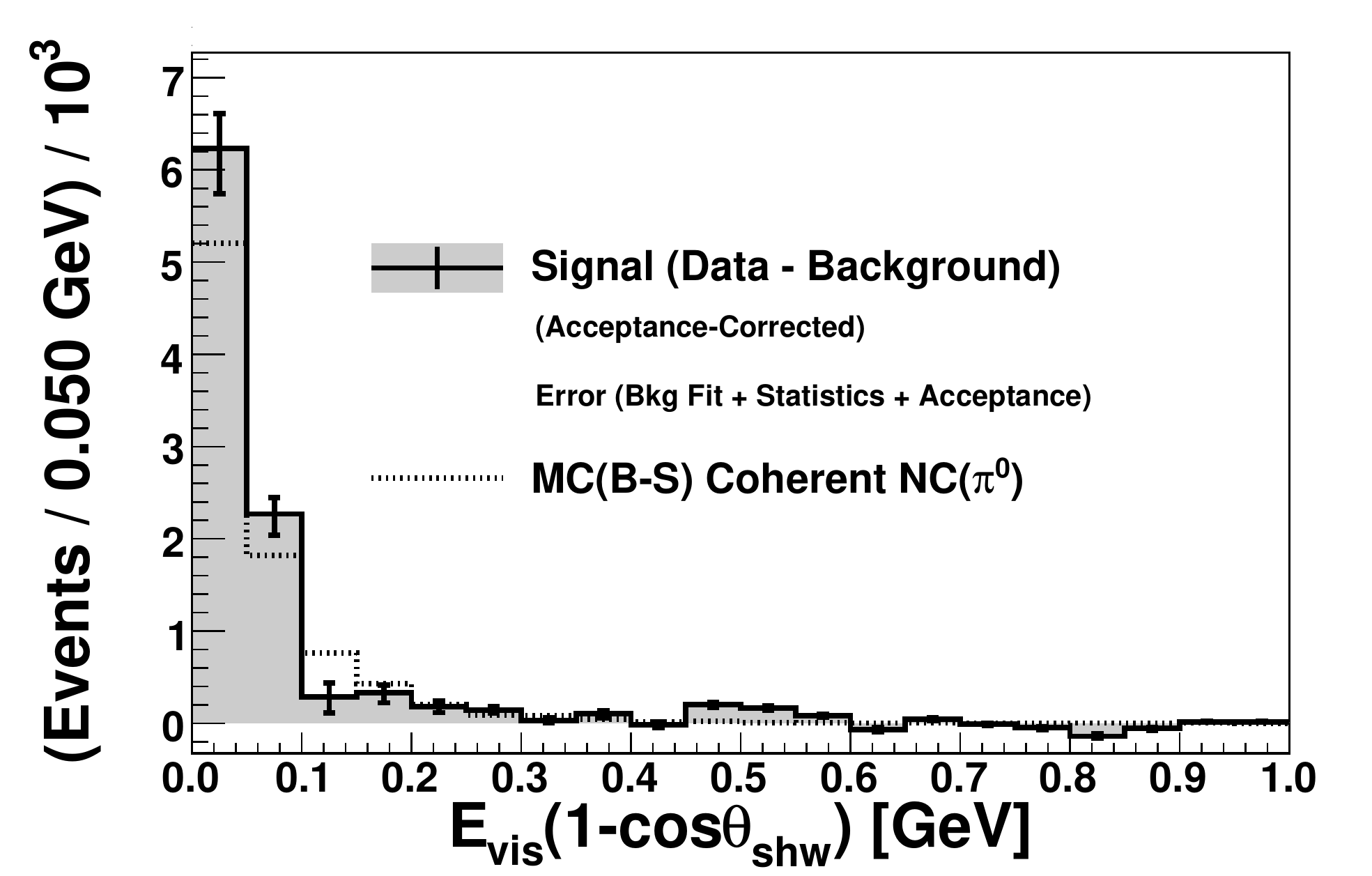}}
 \end{adjustwidth}
 \caption{Distribution in $\etapi$ for the acceptance-corrected data excess above the total neutrino
 background rate, for events having $E_{vis} > 1.0$\,GeV
 (shaded histogram).   The dotted-line histogram shows the Berger-Sehgal prediction.}
 \label{Fig11}
\end{figure}

Signal events are accepted into
the selected sample with an efficiency of $10.7\%$.  (The total acceptance, accounting for loss due
to the $E_{vis} < 1.0$\,GeV threshold cut, is estimated to be $4.6\%$).
The correction to the measured event rate is implemented as prescribed by Eq.\,(\ref{eq:sigExtract_ER}).    
Figure~\ref{Fig11} shows the acceptance-corrected signal as a function of $\etapi$ (shaded
histogram) for all events having $E_{vis} > 1.0$\,GeV.  Error bars on the binned signal are the quadrature sum of background 
uncertainties, statistical errors, and uncertainty with acceptance correction factors.  
In Fig.~\ref{Fig11} the coherent-scattering signal is almost entirely confined to the range $0.0<\etapi< 0.2$ in agreement with
the general trend predicted by the Berger-Sehgal model (dotted-line histogram).    However the data exceed
the model's prediction by nearly 2\,$\sigma$ for $0.0<\etapi<0.1$, while falling below the prediction for $0.1< \etapi<0.2$.
These features suggest that the coherent interaction may be more sharply peaked towards $\etapi = 0$ than is predicted by Berger-Sehgal.

\smallskip

The largest uncertainty of $N^{Coh}$ arises from the estimation and subtraction of background events.  
The constraints on the normalization and systematic parameters obtained from the fit to the sidebands
are the 1$\upsigma$ confidence intervals extracted from the profiled 1-D $\Delta\chi^{2}$ distributions.  
The uncertainty due to the background subtraction is calculated from the minimum and maximum
background rates allowed by the fit-parameter ranges.   Propagation of the 1$\upsigma$ CL interval limits on the 
background to the final measurement results in an error band of (+11.6\%, -14.4\%).

The signal model enters into the calculation of $N^{Coh}$ in the acceptance corrections, and in
the calculation of the signal in the sidebands.    The bin-by-bin acceptance correction factors for 
events with $E_{vis}>1.0$ GeV incur statistical uncertainties from the reference signal MC 
and shape uncertainties due to finite bin widths. 
This contributes an additional $\pm$3.2\% uncertainty and increases the total error band to (+12.0\%, -14.8\%). 
A larger signal-model dependence enters via the correction factor 
used to estimate the number of events having $E_{vis} <1.0$ GeV.   An uncertainty
estimate is presented in  Sec.\,\ref{sec:tot_xsec}.

Biasing of the extracted signal towards the signal-model prediction (5.8\%) is corrected by scaling
the measured signal amount away from the MC prediction by 5.8\%.   The uncertainty  introduced 
by this correction is listed in the second row of Table~\ref{tab:Ncoh_error-total}.
Then the total extracted signal is $N^{Coh}$ = 9,550 events.  
The percentage error range calculated for $N^{Coh}$ at this stage is (+12.6\%/-15.5\%), in good agreement 
with an estimate based upon mock data experiments of $\pm15.8\%$ (see Appendix).
The coherent NC($\uppi^0$) signal is 12.8\% higher than, but within 1$\upsigma$ of, the Berger-Sehgal prediction.

There is uncertainty  in the subtraction of the estimated background from
purely-leptonic neutrino-electron scattering ($\pm0.8\%$).   Additionally
the signal sample may incur a small contribution from 
diffractive scattering of neutrinos on hydrogen~\cite{ref:diffractive-H}.   Based upon a 
calculation by B.\,Z.\,Kopeliovich\,{\it et.\,al.}~\cite{ref:program-NP}, 
the uncertainty introduced by this possible contaminant is estimated to be $< 3.7\%$ of the NC(\piz) signal
predicted by Berger-Sehgal.   These errors are added in quadrature to the error 
on $N^{Coh}$ arising from the fit-based neutrino background subtraction (see rows 4, 5 of 
Table~\ref{tab:Ncoh_error-total}).

Directly applicable to this analysis are evaluations, 
carried out for the MINOS $\nue$ appearance search,  
of uncertainty introduced to EM shower selection 
by uncertainties associated with calibrations~\cite{ref:Boehm_Thesis}.    
Sources of uncertainties include calibrations of photomultiplier
gains, scintillator attenuation, strip-to-strip variation, 
detector non-linearity, and mis-modeling of low pulse height hits.  
The total EM calibration uncertainty is estimated to be $\pm4.7\%$; 
it is added in quadrature to the $N^{Coh}$ determination, 
bringing the cumulative uncertainty on $N^{Coh}$  to $(+13.5\%,-16.6\%)$.

\smallskip
\begin{table}
\centering
\scalebox{1.0}{%
\begin{tabular}{|c|cc|}
\hline
 & \multicolumn{2}{c}{ \rule{0pt}{2.5ex} $N^{Coh}$ $\pm1\upsigma$  Range} \vline \\ 
{\bf Source of Uncertainty}& \multicolumn{2}{c}{ \rule{0pt}{2.1ex} 9550 Evts, $E_{vis} >1.0$\, GeV } \vline \\
\cline{2-3}
     & (+) Shift  & (-) Shift  \\
\cline{1-3}
\cline{1-3}
background subtraction  & ~~~11.6\% & ~14.4\%\\
\cline{1-3}
biasing to signal model  & ~~~~3.8\% & ~~4.6\%\\
\cline{1-3}
acceptance corrections  & ~~~~3.2\% & ~~3.2\%\\
\cline{1-3}
$\nu$ purely leptonic bkgrd & ~~~~0.8\% & ~~0.8\%\\
\cline{1-3}
diffractive scattering ($\nu$H) & ~~~~0.0\% & ~~3.7\%\\
\cline{1-3}
detector EM calibration &    ~~~~4.7\% & ~~4.7\%\\
\cline{1-3}
\hline
\hline
Total Syst. Error  & +13.5\% & -16.6\%\\
\hline
\end{tabular}}
\caption{Composition of the error ($\pm1\upsigma$) on the number of NC($\uppi^0$) coherent scattering events 
($E_{vis} >1.0$\,GeV) determined by the analysis.
\label{tab:Ncoh_error-total}}
\end{table} 

\smallskip
The sensitivity to the $Q^{2}$ dependence of the signal model  was examined using mock data experiments.   
The NC($\uppi^{0}$) coherent scattering content in the sideband samples of mock data experiments 
was varied by amounts representative of plausible changes to the $Q^{2}$ dependence of the signal model.    
The variations were found to introduce negligible changes to the mean value and uncertainty range for
the ensemble of $N^{Coh}$ outcomes of the simulated experiments (see Fig.~\ref{Fig14} of the Appendix).

\section{cross sections}
\label{sec:tot_xsec}

 
A data sample enriched in coherent NC($\uppi^0$) scattering events is now isolated, and an event excess of $5.4\,\upsigma$ 
above the estimated background for this process is observed.
The signal count, $N^{Coh}$, is now converted into a cross section for coherent $\uppi^0$ production with
$E_{vis} > 1.0$\,GeV final states, using Eq.\,(\ref{eq:xsec3}).
The quantities required for the calculation are given in Table~\ref{tab:xsec}.

\begin{figure}
\begin{adjustwidth}{-1.0in}{-1.0in}
 \centering
 \scalebox{0.40}{\includegraphics{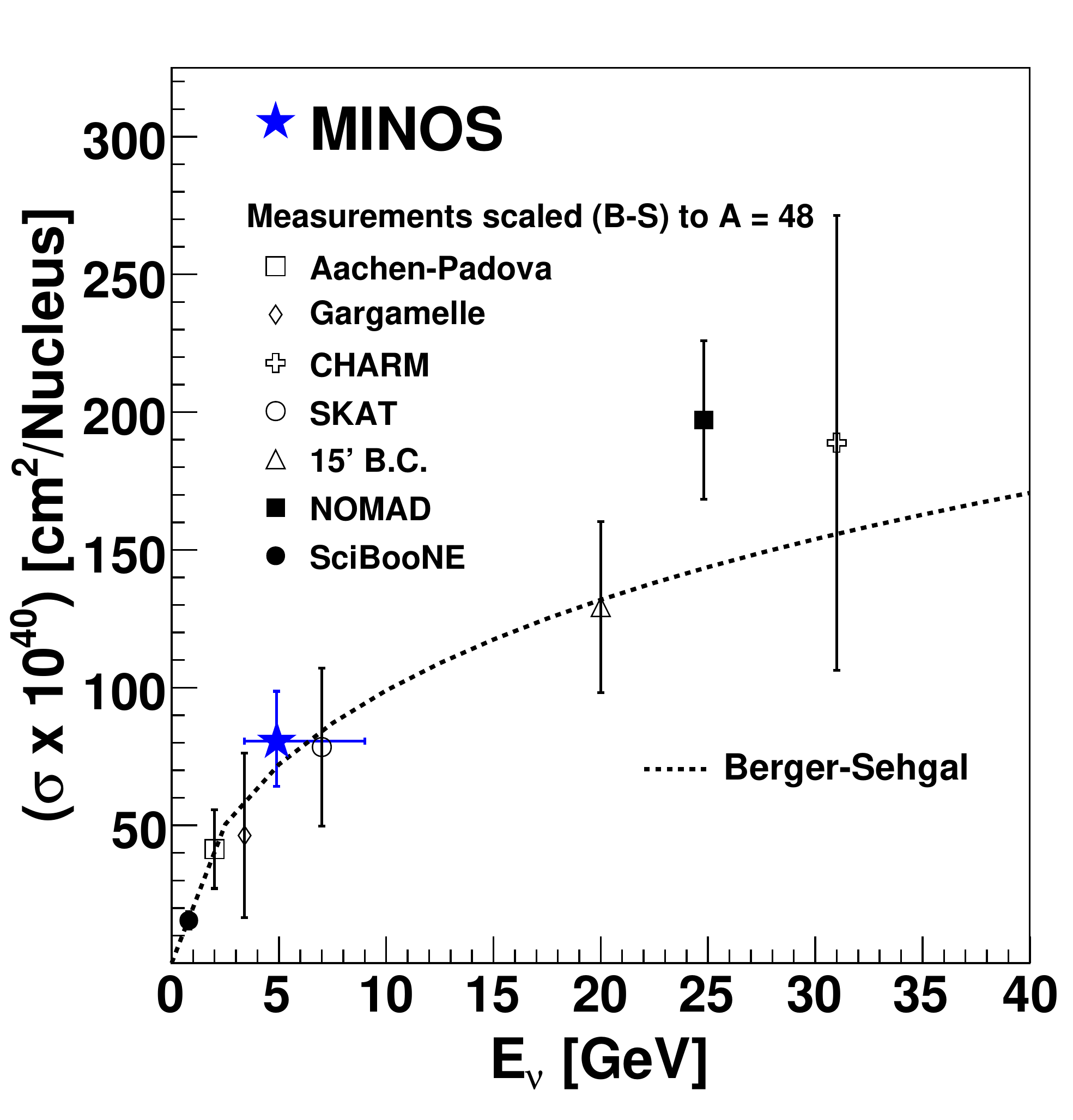}}
 \end{adjustwidth}
 \caption{Comparison of the MINOS NC($\uppi^0$) 
 coherent scattering total cross section (flux-averaged) to measurements
 obtained with $\langle E_{\nu} \rangle$ at lower and higher values.
Previous measurements are shown scaled 
to the MINOS target medium, $\langle \mathcal{A} \rangle = 48$, for the purpose of comparison.
Table~\ref{tab:prev_exp} compares the previous measurements as reported, to the direct
Berger-Sehgal prediction.}
 \label{Fig12}
\end{figure}

The cross section $\langle\upsigma\rangle$ obtained 
is an average over the neutrino flux of the NuMI LE beam for which  the 
average neutrino energy is 4.9\,GeV.
Table~\ref{tab:xsec} shows that the error for $\langle\upsigma\rangle$ is dominated by the total uncertainty
ascribed to the signal extraction (Table~\ref{tab:Ncoh_error-total}),
with the uncertainty for the neutrino flux 
contributing an additional 7.8$\%$.   Inclusion of all sources 
yields a total uncertainty of (+15.6\%, -18.4\%).
The detector medium consists of iron and carbon nuclei with abundances very nearly 80\%:20\%.  
Using Eq.\,(\ref{eq:xsec3}), the flux-averaged, $\mathcal{A}$-averaged coherent scattering cross section for
events above the analysis threshold of $E_{vis} > 1.0\ \text{GeV}$, is
\begin{equation}
\begin{split}
\langle\upsigma\rangle =  32.6\pm2.1\,(\text{stat}) ^{+4.7}_{-5.6}\,(\text{syst})\times10^{-40}~\text{cm}^2/\text{nucleus} \bf{ .} \\
\end{split}
\label{eq:data_xsec1}
\end{equation}
\noindent
In row 1 of Table~\ref{tab:MINOS_xsec}, this result is compared to the flux-averaged cross section predicted by the 
Berger-Sehgal model:  $31 \times 10^{-40}\,\text{cm}^2$ per nucleus.
(The flux-averaged cross sections for an iron:carbon 80$\%$:20$\%$ mixture in Table~\ref{tab:MINOS_xsec} (rows 1 and 4)
are to be distinguished from Berger-Sehgal predictions for a titanium $\mathcal{A}$ = 48 target.
The latter are approximations of the former;   they are used to provide
the dashed curve that serves as a visual aid in Fig.~\ref{Fig12}.)

\begin{table}
\centering
\scalebox{0.80}{%
\begin{tabular}{|c||c|c|c|}
\hline
\bf{Input}                              & \multirow{2}{*}{\bf{Description}}         & \multirow{2}{*}{\bf{Value}}            & \bf{Fractional}         \\
\bf{Parameter}                          &                                           &                                        & \bf{Error}              \\
\hline
\hline
\multirow{2}{*}{${\mathcal{N}_{T}}$} & Number of nuclei in                     & \multirow{2}{*}{3.57$\times$10$^{29}$} & \multirow{2}{*}{0.4\%} \\
                                                         & the fiducial volume                       &                                                               &                         \\
\hline
\multirow{2}{*}{${\mathcal{N}_{\text p}}$}     & Neutrino exposure                 & \multirow{2}{*}{2.8$\times$10$^{20}$}  & \multirow{2}{*}{1.0\%}  \\
                                                             & [POT]                                     &                                                               &                         \\
\hline
\multirow{2}{*}{${\Phi}$}           & Flux                                      & \multirow{2}{*}{2.93$\times$10$^{-8}$} & \multirow{2}{*}{7.8\%}  \\
                                                 & [Neutrinos/POT/cm$^{2}$]   &                                                               &                         \\
\hline
\hline
\multirow{2}{*}{$N^{Coh}$}      & Coherent events (corrected)    & \multirow{2}{*}{9,550}          & \multirow{2}{*}{\Large $^{+13.5\%}_{-16.6\%}$} \\
                                                & ($E_{vis}>1.0$ GeV)                &                                             &                                                                             \\
\hline
\hline
\multirow{2}{*}{$\fint_{\text Fe}$}      & Est. fraction (B-S model)                     & \multirow{2}{*}{0.93}                 & \multirow{2}{*}{1.4\%} \\
                                                 & of coherent events on Fe                     &                                                    &                         \\
\hline
\multirow{2}{*}{${\fint_{\text C}}$}       & Est. fraction (B-S model)                       & \multirow{2}{*}{0.07}                 & \multirow{2}{*}{18.6\%} \\
                                                   & of coherent events on C                         &                                                   &                         \\
\hline
\multirow{2}{*}{$\epsilon^{-1}_{thr}$}& Correction factor for                        & \multirow{2}{*}{2.38}        & \multirow{2}{*}{ $ 13.0\%$} \\
                                                         & $E_{vis}\ge1.0$ GeV threshold       &                                         &                         \\
\hline
\end{tabular}}
\caption{Values and fractional errors for quantities used in cross section determinations based upon Eq.\,(\ref{eq:xsec3}).}
\label{tab:xsec}
\end{table}

The fiducial volume contains $2.89\times 10^{29}$ iron nuclei 
and $6.57\times 10^{28}$ carbon nuclei~\cite{ref:NIM}.
Using these numbers, the coherent scattering cross sections 
on pure iron ($\mathcal{A}$ = 56) versus pure carbon ($\mathcal{A}$ = 12) targets 
can be estimated using the Berger-Sehgal model. 
A 20\% uncertainty is estimated for the iron:carbon cross-section ratio 
based on comparison of Berger-Sehgal with the coherent
scattering calculation of Ref.~\cite{ref:Paschos-2009}
and is propagated to the numbers of events assigned to iron and to carbon
scattering (Table~\ref{tab:MINOS_xsec}, rows 2, 3 and 5, 6).   
The estimated cross sections for iron and for carbon 
scale with the Berger-Sehgal model predictions by construction;  
the uncertainty propagated from the cross-section ratio
covers the model-dependence of these extrapolations.

With the measured partial cross section of Eq.\,\eqref{eq:data_xsec1} in hand, the 
flux-averaged total cross section for Reaction \eqref {eq:nccoh_reaction} 
can now be determined.   Its calculation requires 
a correction factor,  $\epsilon^{-1}_{thr}$, to scale 
the observed event rate to account for loss of signal events whose 
$E_{vis}$ lies below the 1.0\,GeV threshold.    An estimation of
this sizable correction is provided by the Berger-Sehgal based extrapolation 
indicated by Fig.~\ref{Fig03}b: $\epsilon^{-1}_{thr} = 2.38$.   
However, the uncertainty on $\epsilon^{-1}_{thr}$ needs to be ascertained.

\begin{table}
\centering
\scalebox{0.80}{%
\begin{tabular}{|c||c|c|c|c|}
\hline
Target                 & Minimum              & Number of              & MINOS                                    & Berger-Sehgal        \\
Nucleus               & Energy        & Coherent NC($\uppi^0$)    & Cross Section                                     & Cross Section        \\
$\langle \mathcal{A} \rangle$& $E^{min}_{vis}$& Interactions           & per nucleus                        &  per nucleus      \\
\hline
[u]                  & [GeV]                &                        & [10$^{-40}$ cm$^{2}$]                             & [10$^{-40}$ cm$^{2}$] \\
\hline
\hline
\multirow{2}{*}{48}    & \multirow{6}{*}{1.0} & \multirow{2}{*}{~9,550\Large ${^{+1,290}_{-1,590}}$} & \multirow{2}{*}{32.6\Large ${^{+5.1}_{-6.0}}$}   & \multirow{2}{*}{31}\\
                       &                      &                                                      &                                                  &                      \\
\cline{1-1}\cline{3-5}
\multirow{2}{*}{56}    &                      & \multirow{2}{*}{~8,880\Large ${^{+1,210}_{-1,480}}$} & \multirow{2}{*}{37.5\Large ${^{+5.9}_{-6.9}}$}   & \multirow{2}{*}{36}\\
                       &                      &                                                      &                                                  &                      \\
\cline{1-1}\cline{3-5}
\multirow{2}{*}{12}    &                      & \multirow{2}{*}{~~670\Large ${^{\,+\,150}_{\,-\,170}}$} & \multirow{2}{*}{12.4\Large ${^{+3.0}_{-3.2}}$}   & \multirow{2}{*}{11} \\
                       &                      &                                                         &                                                  &                      \\
\hline
\multirow{2}{*}{48}    & \multirow{6}{*}{0.0} & \multirow{2}{*}{22,700\Large ${^{+4,260}_{-4,790}}$} & \multirow{2}{*}{77.6\Large ${^{+15.8}_{-17.5}}$} & \multirow{2}{*}{73}\\
                       &                      &                                                      &                                                  &                      \\
\cline{1-1}\cline{3-5}
\multirow{2}{*}{56}    &                      &\multirow{2}{*}{21,100\Large ${^{+3,970}_{-4,470}}$} & \multirow{2}{*}{89.2\Large ${^{+18.2}_{-20.1}}$} & \multirow{2}{*}{84}\\
                       &                      &                                                     &                                                  &                      \\
\cline{1-1}\cline{3-5}
\multirow{2}{*}{12}    &                      & \multirow{2}{*}{~1,590\Large ${^{+420}_{-470}}$} & \multirow{2}{*}{29.5\Large ${^{+8.1}_{-8.6}}$} & \multirow{2}{*}{29}\\
                       &                      &                                                 &                                                  &                      \\
\hline
\end{tabular}}
\caption{The flux-averaged cross sections $\langle \upsigma \rangle$ for coherent scattering 
in the MINOS medium ($\mathcal{A}$-averaged).   Values for scattering on the component iron and carbon nuclei are
inferred from the $\langle \mathcal{A} \rangle = 48$ measurement. The event rate directly observed determines the partial cross sections  (upper rows).    Correction for rate loss due to 
the threshold cut at $E_{vis} = 1.0$\,GeV yields total cross sections (lower rows).}
\label{tab:MINOS_xsec}
\end{table}
\smallskip

The shape of the $E_{vis}$ distribution predicted by Berger-Sehgal (Fig.~\ref{Fig03}b) 
is very similar to the distribution shapes for 
$E_{\uppi^{+}} > 1.0$\,GeV of $\numu\,CC(\uppi^{+})$  
coherent scattering, and for $E_{\uppi^{-}} > 1.0$\,GeV of 
$\anumu\,CC(\uppi^{-})$  coherent scattering as reported by MINERvA~\cite{ref:CC-Coh-Minerva}.    
A data-driven assignment of uncertainty
for extrapolation of the $E_{vis}$ distribution below 1.0 GeV 
is made possible by the fact that the MINERvA measurements 
used the low-energy NuMI fluxes similar to the one of this work;   moreover
coherent $\numu\,CC(\uppi^{+})$ and 
$\anumu\,CC(\uppi^{-})$ are predicted to have identical cross sections,  
and coherent NC($\uppi^0$) scattering is predicted
to have the same final-state kinematics as coherent CC($\uppi^{\pm}$)~\cite{ref:Paschos}.

With extrapolations of the $E_{vis}$ distribution below the 1.0\,GeV threshold,  
it is found (utilizing the supplemental materials of~\cite{ref:CC-Coh-Minerva}),
that the MINERvA $\numu\,CC(\uppi^{+})$ 
and $\anumu\,CC(\uppi^{-})$ coherent scattering distributions bracket 
the Berger-Sehgal distribution from above and below, respectively.   
The range of plausible alternative shapes for the Berger-Sehgal distribution 
for $E_{vis} < 1.0$ GeV that are compatible with the MINER$\nu$A data, 
implies an uncertainty range for $\epsilon^{-1}_{thr}$.    A complication is that the 
MINER$\nu$A data is coherent scattering on carbon, while the scaling factor illustrated by
Fig.~\ref{Fig03}b is calculated for coherent scattering on iron and carbon, 
so there is uncertainty arising from possible $A$-dependence of the $E_{vis}$ distribution. 
The uncertainty in going from carbon to iron was estimated by comparing to the
Rein-Sehgal model (GENIE implementation), and additionally 
by running the Berger-Sehgal model with variations to input values from pion-nucleus scattering data.
The $A$-dependence uncertainty is found to be the main contributor 
to the uncertainty for $\epsilon^{-1}_{thr}$.   An uncertainty of $\pm 13.0\%$ is assigned, 
as listed in the bottom row of Table~\ref{tab:xsec}.

The total coherent cross section, $\mathcal{A}$-averaged over the MINOS medium and flux-averaged
with $\langle{E_\nu}\rangle$ of 4.9\,GeV, is
\begin{equation}
\begin{split}
\langle\upsigma\rangle =  77.6\pm5.0\,(\text{stat}) ^{+15.0}_{-16.8}\,(\text{syst})\times10^{-40}~\text{cm}^2/\text{nucleus} \bf{ .} 
\label{eq:data_xsec-total}
\end{split}
\end{equation}
The corresponding Berger-Sehgal cross-section prediction is
$73\times10^{-40}\,\text{cm}^2~\text{per nucleus}$.

\section{Discussion and Conclusion}
\label{sec:Results}

 \subsection{Cross section versus $E_{\nu}$ and $\mathcal{A}$}

\begin{figure}
\begin{adjustwidth}{-1.0in}{-1.0in}
 \centering
 \scalebox{0.40}{\includegraphics{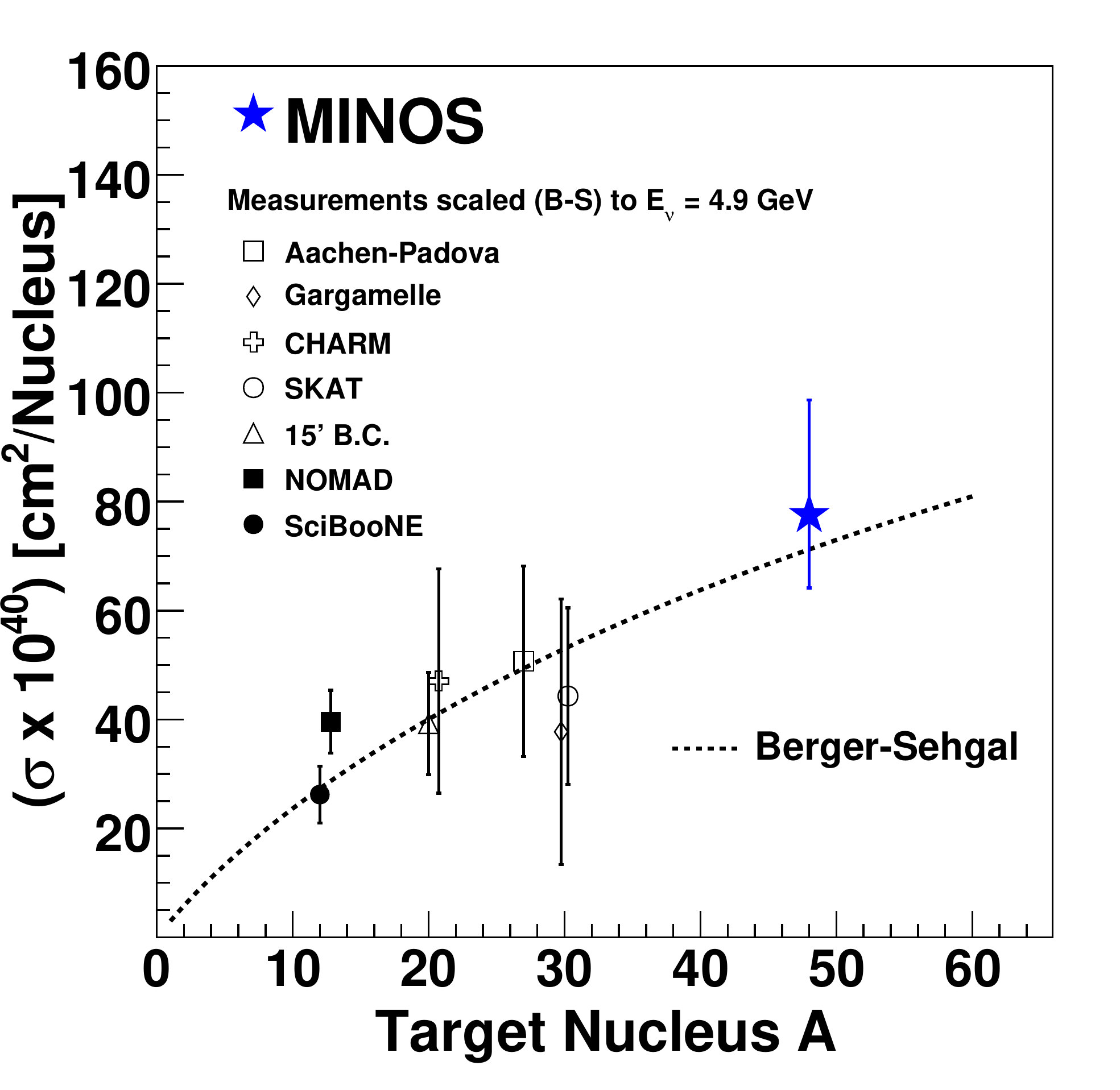}}
 \end{adjustwidth}
\caption{MINOS total cross section (star symbol) 
for neutrino NC($\uppi^0$) coherent scattering at  $\langle{E_\nu}\rangle$ = 4.9\,GeV for
nuclei with $\langle \mathcal{A} \rangle = 48$.   
The previous measurements, listed in Table~\ref{tab:prev_exp}, are shown scaled to the same average neutrino energy.}
\label{Fig13}
\end{figure}

As shown by Table~\ref{tab:prev_exp}, the MINOS measurement examines 
NC($\uppi^0$) coherent scattering in an $E_\nu$ - $\mathcal{A}$ region that lies outside
of the range probed by previous experiments.  
For the purpose of eliciting the $E_{\nu}$ dependence, the
previously reported cross sections (see Table~\ref{tab:prev_exp}) are scaled to an $\mathcal{A}$ = 48 nucleus using the Berger-Sehgal model.   
(The 15-ft Bubble Chamber and SciBooNE cross section measurements are reported as fractions of the 
Rein-Sehgal cross sections for neon~\cite{ref:15BC} and for carbon~\cite{ref:MiniBooNE, ref:SciBooNE}  respectively.)
The scaled cross-section values are plotted in Fig.~\ref{Fig12}  
together with the $\mathcal{A}$-averaged MINOS measurement (solid star).    For purposes of display, the $E_\nu$ interval of the
measurement is taken to be the interval on either side of 4.9\,GeV which includes 34$\%$ of the neutrino flux.
Also shown in Fig.~\ref{Fig12} is the 
prediction for $\mathcal{A}$ = 48 of the Berger-Sehgal model (dashed curve).    
Figures~\ref{Fig12} and \ref{Fig13} show that the ensemble 
of cross-section measurements for Reaction\,\eqref{eq:nccoh_reaction},
when subjected to ``normalization" to common $\langle \mathcal{A} \rangle$ or 
$\langle E_{\nu}\rangle$, exhibit power-law growth with increasing neutrino energy 
for fixed $\mathcal{A}$, or with increasing target nucleon number for fixed $\langle E_{\nu} \rangle$.

The $\mathcal{A}$-dependence of the coherent cross section is examined by
comparing the MINOS result to previous measurements, 
where the latter are scaled to $\langle E_{\nu} \rangle = 4.9$\,GeV
according to the cross-section ratio predicted by Berger-Sehgal.
Figure~\ref{Fig13} compares the measurements obtained for
the different  target $\mathcal{A}$, when their values are scaled in this way.
(The extrapolations via Berger-Sehgal to pure iron and carbon targets
listed in Table~\ref{tab:MINOS_xsec} are not plotted.)
The high-$\langle \mathcal{A} \rangle$ MINOS result is consistent with 
the trend predicted by PCAC models~\cite{ref:Paschos-2009}.   
In rough terms, the $\mathcal{A}$-dependence in the Berger-Sehgal model for $E_{vis} > 1.0$\,GeV arises
from a convolution of three effects.    The coherent nature of the interaction gives
an $\mathcal{A}^2$ dependence,  but that is diminished by the 
nuclear form factor and by pion absorption.  The former falls off as $\exp(-\mathcal{A}^{2/3})$,  
and the latter as $\exp(-\mathcal{A}^{1/3})$.    These effects combine to yield a total cross section 
with an approximate $\mathcal{A}^{2/3}$ dependence.

\subsection{Conclusion}
The MINOS Near Detector is used to study coherent NC production 
of single $\uppi^0$ mesons initiated by neutrino scattering on a target medium 
consisting mostly of iron nuclei, with $\langle \mathcal{A} \rangle = 48$.
Using a low-energy NuMI beam exposure of 2.8$\times 10^{20}$ POT
with mean (mode) $E_{\nu}$ of 4.9 GeV (3.0 GeV), 
a signal sample comprised of 9,550$^{+1,200}_{-1,590}$ events having final-state
$E_{vis} > 1.0$\,GeV has been isolated.    The corresponding flux-averaged, 
$\mathcal{A}$-averaged partial cross section for events above the analysis 
$E_{vis}$ threshold of 1.0 GeV is presented in Eq.~\eqref{eq:data_xsec1}.
Extrapolation of the $E_{vis}$ distribution from the analysis 1.0 GeV threshold to zero yields the total coherent scattering
cross section.    The flux-averaged, $\mathcal{A}$-averaged total cross section is given in Eq.~\eqref{eq:data_xsec-total}.
Its value is $\langle\upsigma\rangle = (77.6^{+15.8}_{-17.5})\times10^{-40}~\text{cm}^2~\text{per nucleus}$.  
The various neutrino-nucleus NC($\uppi^0$) coherent scattering cross sections that are measured or inferred from this
work are listed in Table~\ref{tab:MINOS_xsec}.   The measurements of coherent scattering Reaction\,\eqref{eq:nccoh_reaction}
reported here are the first to utilize a target medium
of average nucleon number $\langle \mathcal{A} \rangle > 30$, and the cross section results 
of Eqs.~\eqref{eq:data_xsec1} and \eqref{eq:data_xsec-total} are for coherent scattering at the highest average nucleon number
obtained by any experiment to date.
Figures \ref{Fig12} and \ref{Fig13} show that these cross sections, as with
previous measurements on lighter
nuclear media and at lower and higher $\langle{E_\nu}\rangle$ values, exhibit the general trends
predicted by the Berger-Sehgal coherent scattering model which is founded upon
PCAC phenomenology.

\vspace{+7pt}
\section{Acknowledgments} \vspace{-8pt}

This work was supported by the US DOE; the UK STFC; the US NSF; the State and University of Minnesota; the
University of Athens, Greece; and Brazil's FAPESP, CNPq, and CAPES.
We gratefully acknowledge the staff of Fermilab for invaluable contributions to the research reported here.

\section*{Appendix: Fit Validation}
\label{sec:SimExp}

\smallskip
Realistic mock data experiments were used to 
validate the analysis fitting procedure~\cite{ref:Cherdack-Thesis}.
The generation of mock data for the latter simulated experiments is more elaborate than for the SSMD experiments.   
As with SSMD experiments, each mock data sample provides a population of events extending 
over the $\cos\theta_{shw}$-vs-$E_{vis}$ plane, binned in the same way as for the observed data. 
However with the full mock data samples,  the background templates are adjusted to reflect random fluctuations in
each systematic parameter, and the coherent signal content was varied by adjusting the normalization of the 
signal model over the range $\pm50\%$.
Statistical fluctuations are applied to the event totals in each bin after the templates are combined. 
The entire background fitting and signal extraction procedure 
is executed on an ensemble of these mock data samples.
Each mock data ``experiment" yields a set of best-fit values for the fit parameters, a best-fit $\chi^{2}$, and an
acceptance-corrected event rate $N_{fit}$, to be compared
to the ``true" signal assumed for the 
simulated experiment, $N_{input}$.

\begin{figure}
\begin{adjustwidth}{-1.0in}{-1.0in}
 \centering
 \scalebox{0.43}{\includegraphics[angle=0]{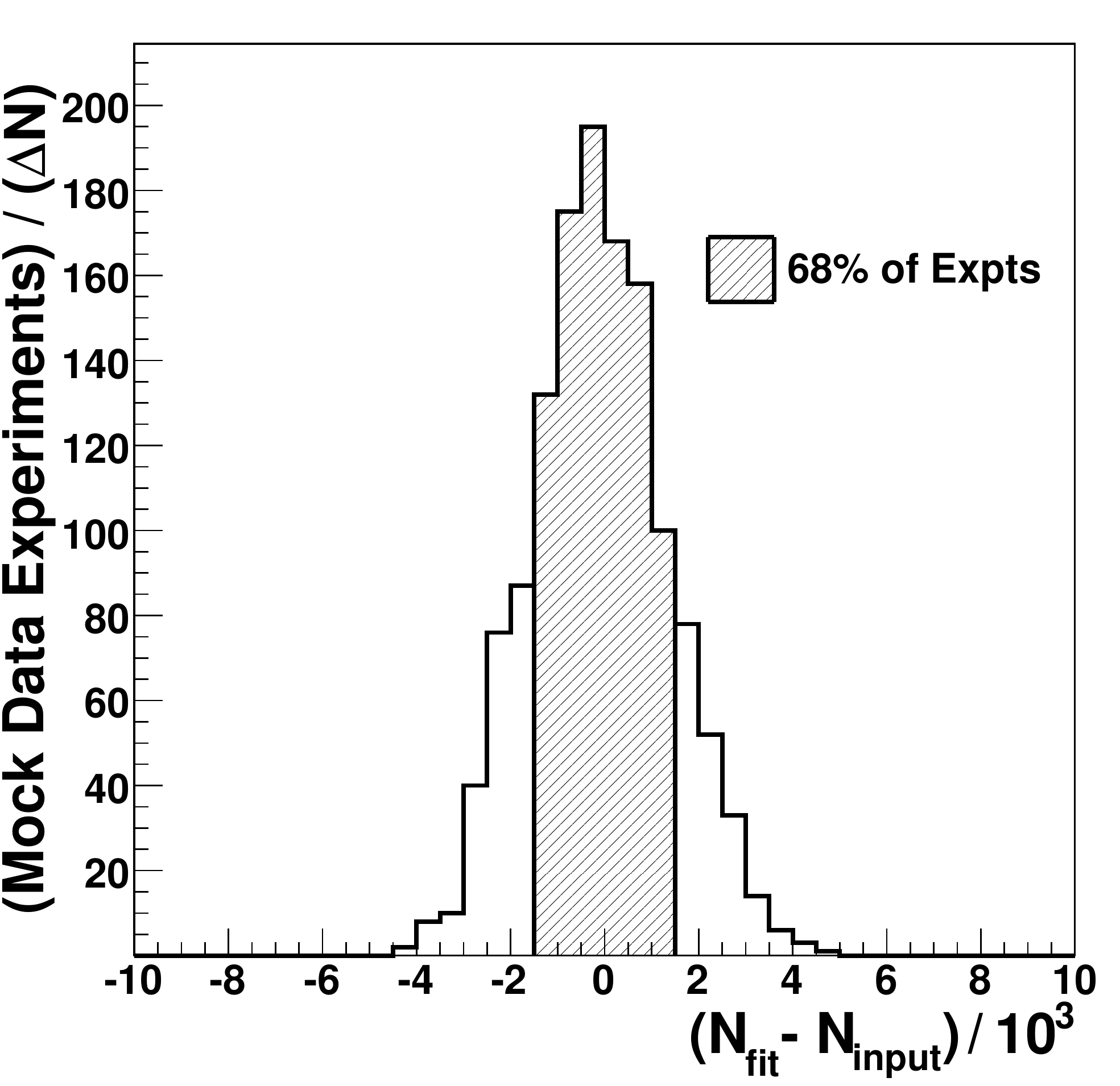}}          
 \end{adjustwidth}
 \caption{Deviation of best-fit outcomes $N_{fit}$                                          
 from coherent signal inputs $N_{input}$, for mock
 data experiments. }
 \label{Fig14}
\end{figure}

Figure~\ref{Fig14} shows the
distribution of $N_{fit} - N_{input}$ for an ensemble 
of mock data experiments.   
The 1$\upsigma$ width, 
defined as the region about the peak that includes $68\%$ of the area, 
was shown to be independent of the input signal normalization.
This metric serves as an estimate of the $\pm 1\upsigma$ confidence interval 
for the final fit procedure, and is measured to be $\pm15.8\%$.
This estimate serves as a cross-check of the uncertainties 
on the measured signal event rate derived from fitting the data.


\end{document}